\documentclass[aps, pre, reprint, amsmath, amssymb, superscriptaddress]{revtex4-2}

\usepackage{amsmath,amssymb,bm, xcolor}
\usepackage{braket}
\usepackage{graphicx}
\graphicspath{{./figures}}
\usepackage{dcolumn}
\usepackage{bm}
\usepackage{lipsum}
\usepackage[colorlinks=true,linkcolor=blue,urlcolor=blue,citecolor=blue]{hyperref}
\usepackage[italicdiff]{physics}
\usepackage{bbold}
\usepackage{dsfont}

\begin{document}

\title{Thermodynamic universality across dissipative quantum phase transitions}

\author{Laetitia P. Bettmann}
\email{bettmanl@tcd.ie}
\affiliation{School of Physics, Trinity College Dublin, College Green, Dublin 2, D02K8N4, Ireland}

\author{Artur M. Lacerda}
\email{machadoa@tcd.ie}
\affiliation{School of Physics, Trinity College Dublin, College Green, Dublin 2, D02K8N4, Ireland}

\author{Mark T. Mitchison}
\email{mark.mitchison@kcl.ac.uk}
\affiliation{Department of Physics, King’s College London, Strand, London, WC2R 2LS, United Kingdom}

\author{John Goold}
\email{gooldj@tcd.ie}
\affiliation{School of Physics, Trinity College Dublin, College Green, Dublin 2, D02K8N4, Ireland}
\affiliation{Trinity Quantum Alliance, Unit 16, Trinity Technology and Enterprise Centre, Pearse Street, Dublin 2, D02YN67, Ireland}

\begin{abstract}
We study finite-time driving across second-order dissipative quantum phase transitions described by Lindblad dynamics. We show that the nonadiabatic entropy production, which quantifies deviations from the instantaneous nonequilibrium steady state, exhibits universal power-law scaling with the ramp duration in analogy to the Kibble--Zurek mechanism for closed systems. This establishes the universality of irreversible dissipation induced by driving an open quantum system near criticality. Furthermore, in systems described by bosonic Gaussian states, our scaling laws predict that the nonadiabatic entropy production is independent of driving speed to leading order, revealing a distinctive feature of Gaussian dissipative quantum phase transitions.
We validate these analytical predictions in the thermodynamic limit of the driven-dissipative Dicke model and via finite-size scaling in the open Kerr model. Our results establish a general framework for understanding universal nonequilibrium response and thermodynamic irreversibility in critical open quantum systems.

\end{abstract}

\maketitle
{\bf Introduction}.---Phase transitions are characterized by diverging relaxation times and correlation lengths, giving rise to universal behavior near critical points~\cite{Kibble1976, Zurek1985, polkovnikov_rmp_2011, delCampo2014, Dziarmaga2010, biroli_kibble-zurek_2010}. In equilibrium quantum systems, continuous quantum phase transitions (QPTs)~\cite{Sachdev2011} occur at zero temperature when ground-state observables or their derivatives show nonanalytic dependence on a control parameter $g$. The energy gap closes at the critical point as $\Delta_H(g) \sim |g-g_c|^{z\nu}$, where $z$ and $\nu$ are the dynamical and correlation length exponents, respectively, and the relaxation time diverges as $\tau_R \sim \Delta_H^{-1}$~\cite{zurek_dynamics_2005}, underpinning the phenomenon of critical slowing down.

Finite-rate driving protocols inevitably generate nonequilibrium excitations. If the driving is slow relative to $\tau_R$, the adiabatic theorem guarantees that the system remains close to its ground state. Near a critical point, however, the diverging relaxation time ensures a breakdown of adiabaticity, captured by the Kibble--Zurek mechanism (KZM)~\cite{Kibble1976, Zurek1985,Dziarmaga2010, delCampo2014}. For a protocol of duration $\tau_q$, the KZM predicts that the number of excitations created is a universal function of $\tau_q$, determined by the critical exponents $z$ and $\nu$~\cite{zurek_dynamics_2005,Polkovnikov2005}. Cumulants of the dissipated work, i.e.~the work performed in excess of the equilibrium free energy difference, also show universal scaling with $\tau_q$~\cite{deffner_kibble-zurek_2017,fei_work_2020,naze_kibblezurek_2022,Ma2025, ShenQuantifyingMinimizing2025}. This universality of thermodynamic dissipation reflects the fact that finite-speed driving across a critical point is an inherently irreversible process.

Even richer phenomenology arises in open quantum systems, which may undergo phase transitions between nonequilibrium steady states (NESSs) upon variation of a control parameter~\cite{Sieberer_universality_2025}. These \textit{dissipative} QPTs (DPQTs) have been observed in systems ranging from ultracold atomic gases~\cite{Baumann2010, Ferri2021} to solid-state optical~\cite{fitzpatrick_observation_2017,rodriguez_probing_2017,fink_signatures_2018} and electronic~\cite{beaulieu_observation_2025} devices, allowing for time-resolved measurements of critical fluctuations~\cite{Brennecke2013} and entropy production~\cite{brunelli_experimental_2018} as a DQPT is traversed. Signatures of KZM-like physics have also been observed in classical~\cite{ducci_order_1999, casado_birth_2007, casado_testing_2006, casado_topological_2001} and quantum~\cite{klinder_dynamical_2015,zamora_kibble-zurek_2020,hedvall_dynamics_2017} nonequilibrium phase transitions. However, since a NESS continuously produces entropy even in the absence of time-dependent forcing, the universal thermodynamic properties of dissipation induced by driving across a DQPT have, so far, remained elusive. 

In this work, we use recent insights from information geometry~\cite{lacerda_information_2025} to solve this open problem. 
We show that a general measure of excess dissipation---the nonadiabatic entropy production---scales universally with the protocol duration $\tau_q$ when driving across a DQPT, thus constituting a genuinely far-from-equilibrium analogue of the KZM. 
We focus on Markovian dynamics described by a Lindblad master equation $\mathrm{d}\hat\rho/\mathrm{d}t = \mathcal{L}_g \hat\rho$, with a Liouvillian superoperator $\mathcal{L}_g$ depending on a control parameter $g$ and a NESS $\hat{\pi}_g$ satisfying $\mathcal{L}_g \hat{\pi}_g = 0$. Relaxation towards the NESS is controlled by the Liouvillian gap $\Delta_g := \min_{\lambda_i\neq 0} \mathrm{Re}[-\lambda_i]$, with $\{\lambda_i\}$ the Liouvillian eigenvalues. Near a second-order DQPT at $g=g_c$, the gap closes algebraically~\cite{verstraelen_classical_2020, zhang_driven-dissipative_2021, kessler_dissipative_2012, torre_keldysh_2013},
\begin{equation}
\label{eq:gap_closing}
\Delta_g \sim |g-g_c|^\gamma,
\end{equation}
setting a diverging relaxation time $\tau_R = \Delta_g^{-1}$, with $\Delta_g$ and $\gamma$ replacing $\Delta_H$ and $z\nu$ of equilibrium QPTs, respectively.
In general, one can separate the rate of change of the von Neumann entropy as
\begin{equation}
\dot S_\mathrm{vN}(t) = \dot\Sigma(t) - \dot\Pi(t),
\end{equation}
with $\dot\Sigma \geq 0$ the entropy production rate and $\dot\Pi$ the entropy flux to the environment. In a driven system, the entropy production rate further decomposes as~\cite{hatano_steadystate_2001, Esposito2010dft, horowitz_equivalent_2014, manzano_quantum_2018}
\begin{equation}
\dot\Sigma(t) = \dot\Sigma_\mathrm{ad}(t) + \dot\Sigma_\mathrm{na}(t).
\end{equation}
Here, $\dot\Sigma_\mathrm{ad}$ is the adiabatic entropy production rate, setting the baseline dissipation needed to maintain the NESS for fixed $g$, and $\dot\Sigma_\mathrm{na}$ is the nonadiabatic entropy production rate, quantifying the additional irreversibility due to relaxation of Liouvillian eigenmodes excited by the drive. Crucially, unlike in closed systems, there is no natural notion of a ``number of excitations'' in open dynamics: only the NESS is a normalized density operator, whereas all other Liouvillian eigenmodes are traceless decay modes rather than states. Their ``occupation'' (i.e. spectral weight) therefore has no interpretation in terms of excitations above a ground state, which motivates $\dot\Sigma_\mathrm{na}$ as a universal, thermodynamic quantifier of irreversibility when driving across a DQPT. In the slow-driving regime, $\dot\Sigma_\mathrm{na}$ can be related to the Kubo-Mori-Bogoliubov quantum Fisher information (QFI) metric~\cite{lacerda_information_2025, mandal_analysis_2016}, which describes the susceptibility of the NESS to variations of $g$.

 Our central result is that the total nonadiabatic entropy production $\Sigma_{\rm na}(0,\tau_q) = \int_0^{\tau_q} \mathrm{d}t \,\dot{\Sigma}_{\rm na}$ exhibits universal power-law scaling with the ramp duration $\tau_q$, governed by the Liouvillian gap exponent $\gamma$ and the QFI critical exponent $\alpha$:
\begin{equation}
\label{eq:main_result}
\Sigma_\mathrm{na}(0,\tau_q) \sim \tau_q^{\frac{\alpha-2}{\gamma+1}}.
\end{equation}
Eq.~\eqref{eq:main_result} establishes universality of the excess dissipation induced by finite-speed driving across second-order DQPTs. Universal scaling is also found for the rate $\dot{\Sigma}_{\rm na}(t^*)$ at the freeze-out time $t^*$, where adiabaticity breaks down. Furthermore, we show that $\alpha=2$ for systems in the mean-field Ising universality class, which can be described by Gaussian bosonic states near criticality. Remarkably, in this case, $\Sigma_{\rm na}$ becomes independent of the driving speed, indicating a complete breakdown of adiabaticity even in the limit of infinitely slow driving. This behavior contrasts starkly with the KZM in near-equilibrium systems~\cite{Polkovnikov2005,Acevedo2014, hwang_quantum_2015}, highlighting a distinctive universality of nonequilibrium Gaussian bosonic criticality. We validate these predictions with the examples of the driven-dissipative Dicke and Kerr models.
\begin{figure}
    \centering
\includegraphics[width=\linewidth]{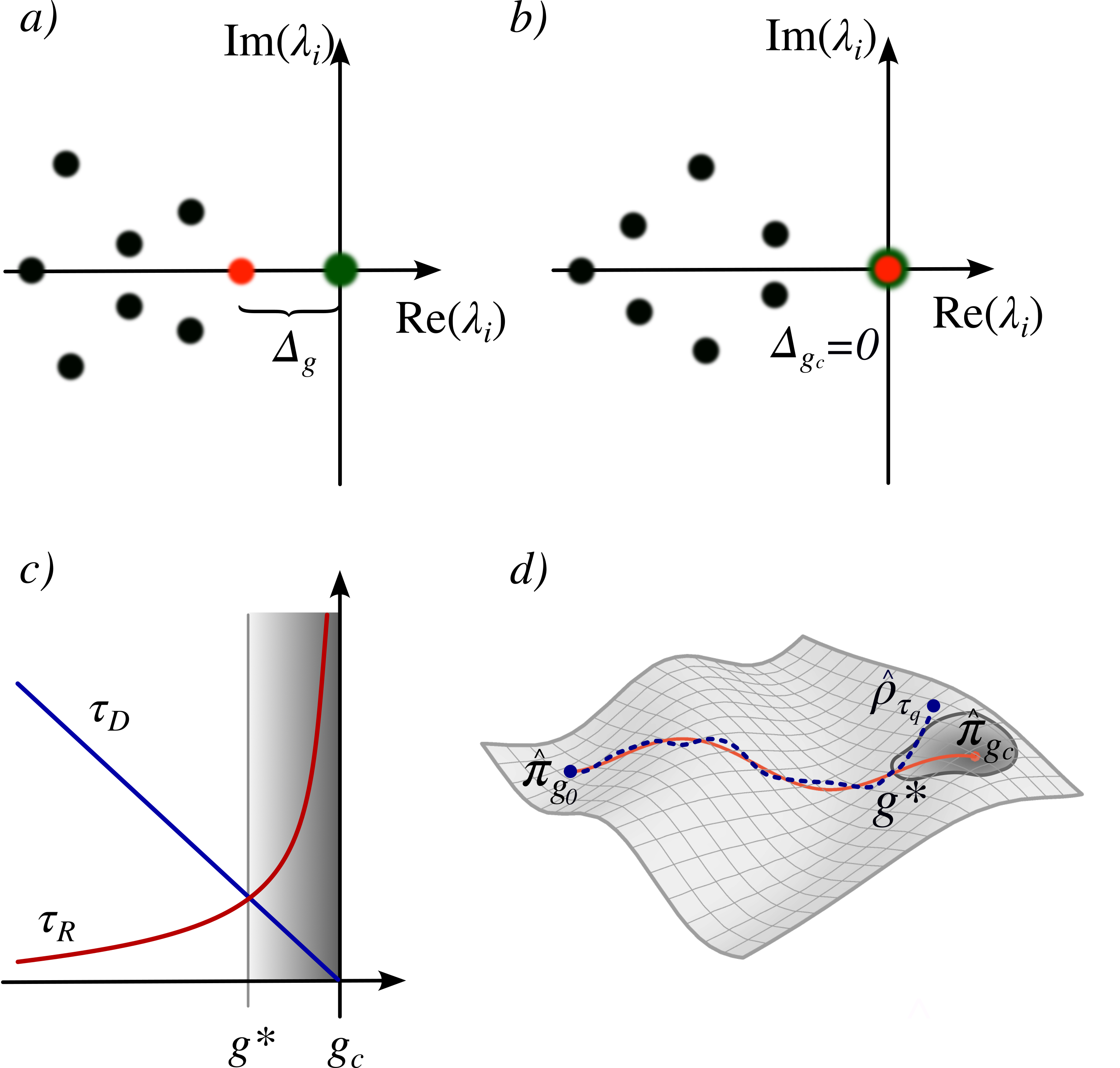}
    \caption{a)  The Liouvillian spectrum $\{\lambda_i\}$, shown in the complex plane, has a finite Liouvillian gap $\Delta_g$ away from the critical point. The soft-mode eigenvalue is shown in red, and the zero eigenvalue corresponding to the unique steady state $\hat{\pi}_g$ in green. b) At the critical coupling $g_c$, the Liouvillian gap closes.  c), d) During a slow, finite-time protocol approaching criticality, the evolving state (blue line in d)) eventually fails to adiabatically follow the instantaneous steady state (red line in d)) and departs from the steady-state manifold. This defines the freeze-out point, $g^* = g(t^*)$, occurring when the intrinsic relaxation time $\tau_R$ exceeds the driving timescale $\tau_D$~\eqref{eq:freeze:out:condition}. The protocol thus separates into a quasiadiabatic and an impulse regime, indicated in grey.
 }
    \label{fig:KZM_OQS}
\end{figure}

{\bf Setup}.---We consider linear ramps of a control parameter towards a DQPT, 
\begin{equation}
g(t) = g_0 + (g_c - g_0)\frac{t}{\tau_q},
\quad t\in [0,\tau_q],
\end{equation}
with quench duration $\tau_q$. So long as the driving timescale $\tau_D(g) = |{\Delta_{g}/\dot \Delta_{g}}|$ is much longer than the relaxation time $\tau_R(g) = \Delta_g^{-1}$, the system remains close to the instantaneous steady state. As $g\!\to\! g_c$, however, $\tau_R$ diverges according to Eq.~\eqref{eq:gap_closing} and the system enters the impulse regime where adiabaticity breaks down. In analogy to the analysis of quench dynamics near QPTs in closed systems~\cite{zurek_dynamics_2005, delCampo2014, de_grandi_quench_2010}, we define the onset of the impulse regime as the freeze-out time ${t}^*$ where the driving timescale matches the relaxation time (see Fig.~\ref{fig:KZM_OQS}), so that
\begin{equation}
\label{eq:freeze:out:condition}
|\dot{\Delta}_{g^*}| = \Delta_{g^*}^2, \qquad g^*=g(t^*).
\end{equation}
The impulse duration $\bar{t} = \tau_q - t^*$ and the distance to criticality then scale as
\begin{equation}
\label{eq:KZ:freezeout}
\bar t \sim \tau_q^{\frac{\gamma}{1+\gamma}}, \qquad |g^*-g_c|\sim \tau_q^{-\frac{1}{1+\gamma}}.
\end{equation}
These results mirror the KZM in closed systems~\cite{delCampo2014, zurek_dynamics_2005, de_grandi_quench_2010, Polkovnikov2005, Barankov_optimal_2008}, with the Liouvillian gap replacing the energy gap.

{\bf Universality of dissipation.}---In driven open quantum systems, finite driving speeds cause deviations from instantaneous steady states. A key thermodynamic quantity characterizing these deviations is the nonadiabatic entropy production rate $\dot{\Sigma}_{\mathrm{na}}$, which can be expressed as~\cite{hatano_steadystate_2001, Esposito2010dft, horowitz_equivalent_2014, manzano_quantum_2018}
\begin{equation}
\label{eq:sigma:na:dot}
    \dot \Sigma_{\mathrm{na}}(t) = - \frac{\mathrm d}{\mathrm ds}D(\hat\rho_s\Vert\hat\pi_{g(t)})\biggr|_{s=t},
\end{equation}
where $D(\hat\rho\Vert\hat\sigma)$ is the quantum relative entropy, $\hat{\rho}_t$ denotes the time-evolving system state and $\hat\pi_{g(t)}$ is the instantaneous steady state that satisfies $\mathcal L_{g(t)} \hat\pi_{g(t)} = 0$.  If $\hat \rho_t$ remains close to $\hat\pi_{g(t)}$, as is typically the case if the driving speed is sufficiently slow, $\hat \rho_t$ can be expanded perturbatively around $\hat\pi_{g(t)}$ in powers of the driving speed $\dot g\sim 1/\tau_q$~\cite{cavina_slow_2017,lacerda_information_2025, mandal_analysis_2016, sivak_thermodynamic_2012, crooks_entropy_1999, scandi_thermodynamic_2019}.
As was recently shown in Ref.~\cite{lacerda_information_2025} (see also Appendix~\ref{app:thermo:metric} for details), the leading nonvanishing contribution to $\dot \Sigma_{\mathrm{na}}(t)$ in such an expansion is quadratric in $\dot g$ and takes the geometric form
\begin{equation}
\label{eq:nonadiabatic:ep:metric:rate}
\dot\Sigma_{\mathrm{na}}(t) \approx  \dot g(t)\, \zeta_{g(t)} \, \dot g(t),
\end{equation}
where the metric, $\zeta_g = \tau_g\, \mathcal{I}^\mathrm{KMB}_g$, factorizes into a generalized integral relaxation time $\tau_g$ from linear response~\cite{kubo_statistical_2012} and the Kubo-Mori-Bogoliubov QFI $\mathcal{I}^\mathrm{KMB}_g$ of $\hat\pi_g$ with respect to the driven parameter $g$. Importantly, the assumptions required for the validity of Eq.~\eqref{eq:nonadiabatic:ep:metric:rate} hold up to the freeze-out point in a slow finite-time ramp, as demonstrated in Fig.~\ref{fig:ODM_Sigmana_scaling} a).

In Appendix~\ref{app:tau_gap}, we show that near a dissipative critical point, the integral relaxation time is inversely proportional to the Liouvillian gap, $\tau_g \sim \Delta_g^{-1}$. Furthermore, since the QFI quantifies the steady state's susceptibility to changes in $g$, it is also expected to diverge algebraically near a second-order DQPT, $\mathcal{I}^\mathrm{KMB}_g \sim |g-g_c|^{-\alpha}$, so that
\begin{equation}
    \label{eq:zeta}
    \zeta_g \sim |g-g_c|^{-(\alpha+\gamma)}.
\end{equation}
With this we arrive at our first main result: for a linear ramp, evaluating Eq.~\eqref{eq:sigma:na:dot} at the freeze-out point~\eqref{eq:KZ:freezeout} yields
\begin{equation}
    \dot\Sigma_{\mathrm{na}}(t^*) \sim \tau_q^{\frac{\alpha-2-\gamma}{\gamma+1}}.
\end{equation}

While the above describes the rate of dissipation at the onset of the impulse regime $t^*$, we now turn to the total nonadiabatic entropy produced over a protocol of duration $\tau_q$. Note that related discussions of total entropy production in scaling contexts can be found in studies of classical phase transitions~\cite{deffner_kibble-zurek_2017} and closed-system QPTs~\cite{naze_kibblezurek_2022}. We obtain the total nonadiabatic entropy production by integrating $\dot{\Sigma}_{\rm na}$ over both the quasiadiabatic and impulse regions as
\begin{equation}
\label{eq:na_quasi_impulse}
\begin{split}
    \Sigma_{\mathrm{na}}(0,\tau_q) &= \int_0^{t^*}\mathrm{d}t\, \dot\Sigma_{\mathrm{na}}(t) + \int_{t^*}^{\tau_q}\mathrm{d}t\, \dot\Sigma_{\mathrm{na}}(t) \\
    &= \Sigma_{\mathrm{na}}(0,t^*)  + \Sigma_{\mathrm{na}}(t^*,\tau_q) .
\end{split}
\end{equation}
Notice that Eq.~\eqref{eq:nonadiabatic:ep:metric:rate} relies on the assumption that the system's state remains close to the corresponding instantaneous steady state. Crucially, this approximation is valid only up to the freeze-out point; beyond this, in the impulse regime, it no longer holds (see Fig.~\ref{fig:ODM_Sigmana_scaling} a)). Therefore, the two regimes must be treated on distinct footings.

Up to the freeze-out point, we compute the nonadiabatic entropy production as the path action with respect to the thermodynamic metric~\cite{lacerda_information_2025}
\begin{equation}
\label{eq:nonadiabatic:ep:metric}
\Sigma_{\mathrm{na}}(0,t^*) = \int_0^{t^*}\mathrm{d}t\, \dot g(t)\, \zeta_{g(t)}\, \dot g(t).
\end{equation}
Employing the freeze-out predictions~\eqref{eq:KZ:freezeout}, we find
\begin{align}
\label{eq:freeze_out_na_total}
\Sigma_{\mathrm{na}}(0,t^*) &\sim \frac{1}{\tau_q} \int_{g_0}^{g(t^*)} \mathrm{d}g\, |g-g_c|^{-(\gamma+\alpha)} \notag \\
&\sim 
\begin{cases}
\frac{1}{\tau_q} \ln \tau_q, & \gamma + \alpha = 1,\\
\tau_q^{\frac{\alpha-2}{\gamma+1}}, & \text{otherwise.}
\end{cases}
\end{align}
Conversely, in the impulse regime ($t\in(t^*,\tau_q]$), critical slowing down implies that the state hardly evolves over the time taken for $g$ to reach $g_c$. We can therefore approximate $\mathcal{L}_{g(t)} \approx \mathcal{L}_{g_c}$ and model the dynamics as relaxation under the final generator with steady state $\hat{\pi}_{g_c}$~\footnote{To be explicit, we use Eq.~\eqref{eq:sigma:na:dot} under the approximation that $\hat{\rho}_t \approx e^{\mathcal{L}_{g_c}(t-t^*)}\hat{\rho}_{t^*}$ and $\hat{\pi}_{g(t)} \approx \hat{\pi}_{g_c}$. This is the standard formulation of entropy production for an evolution with fixed Lindblad generator $\mathcal{L}_{g_c}$~\cite{spohnEntropyProductionQuantum1978}. Here $\hat{\pi}_{g_c}$ should be interpreted in the limiting sense as $g_c$ is approached from below, since the steady state is, strictly speaking, ill-defined in the thermodynamic limit at $g = g_c$.}. This approximation holds in the slow-driving limit of large $\tau_q$, where $\delta g = |g^*-g_c|$ can be treated as a small parameter. The integrated nonadiabatic entropy production then reads
\begin{equation}
\label{eq:sigma:drop:RE}
\Sigma_{\mathrm{na}}(t^*,\tau_q) = D\big(\hat\rho_{t^*}\Vert \hat\pi_{g_c}\big) - D\big(\hat\rho_{\tau_q}\Vert \hat\pi_{g_c}\big).
\end{equation}
This is a positive quantity given by the difference of two positive terms. Therefore, the second term is either subleading with respect to the first or has the same scaling behaviour in the large $\tau_q$ limit. It follows that the scaling of $\Sigma_{\mathrm{na}}(t^*,\tau_q)$ is determined by the behaviour of the first term: either this dominates the second term or the two terms cancel at leading order, leaving a subleading correction. Focusing thus on the first term in Eq.~\eqref{eq:sigma:drop:RE}, we use that $\hat\rho_{t^*}\approx \hat\pi_{g^*}$ in the slow-driving limit, and so
\begin{equation}
\Sigma_{\mathrm{na}}(t^*,\tau_q) \sim D\big(\hat\pi_{g^*}\Vert \hat\pi_{g_c}\big)= \frac{1}{2} (\delta g)^2\, \mathcal I^{\mathrm{KMB}}_{g^*} + O(\delta g^3),
\end{equation}
by the definition of the QFI.
It follows that
\begin{equation}
\Sigma_{\mathrm{na}}(t^*,\tau_q) \sim |g^*-g_c|^{2-\alpha},
\end{equation}
and, using the standard Kibble--Zurek scaling of $\delta g$ with the quench time, we obtain
\begin{equation}
\Sigma_{\mathrm{na}}(t^*,\tau_q) \sim \tau_q^{\frac{\alpha-2}{\gamma+1}}.
\end{equation}
This mirrors the scaling behavior in the quasiadiabatic regime if $\gamma + \alpha \neq 1$; otherwise, it is subleading compared to Eq.~\eqref{eq:freeze_out_na_total}. We thus conclude that, in general, the critical scaling of $\Sigma_{\rm na}(0,\tau_q)$ is governed by Eq.~\eqref{eq:freeze_out_na_total}, which recovers our main result~\eqref{eq:main_result}.
\begin{figure*}
    \centering
\includegraphics[width=\linewidth]{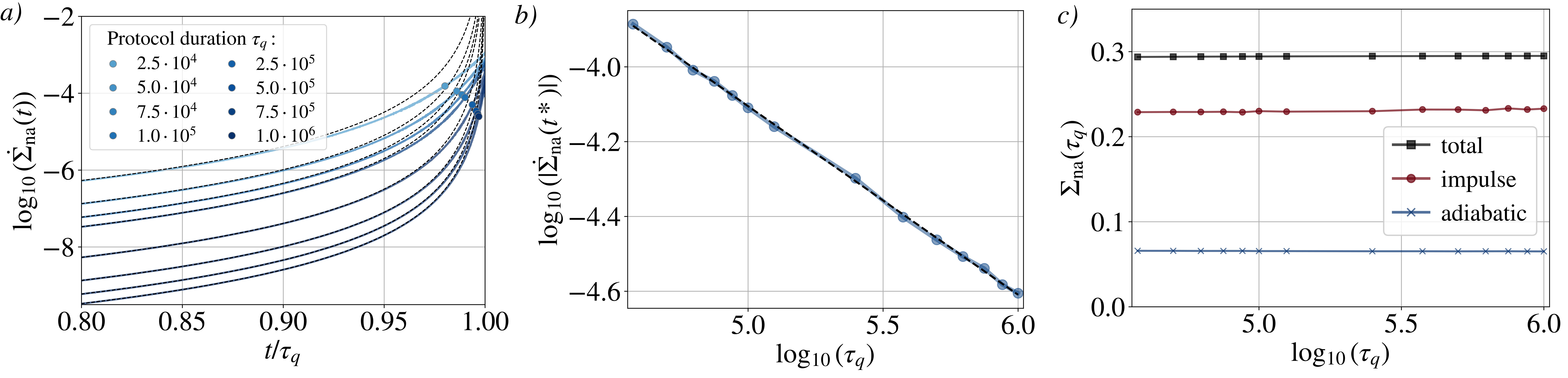}
    \caption{Example of a dissipative quantum phase transition (DQPT) in the open Dicke model. a) The nonadiabatic entropy production rate is shown as a function of time for linear ramps of the coupling strength $g$ from the normal phase to the critical point $g_c$, with protocol durations $\tau_q$ specified in the legend. Scatter points mark their respective freeze-out points $t^*$, set by~\eqref{eq:freeze:out:condition}, solid lines show the exact result~\eqref{eq:sigma:na:dot}, and black dashed lines correspond to the slow-driving approximation~\eqref{eq:nonadiabatic:ep:metric:rate}.  b) Scaling of the nonadiabatic entropy production rate at the onset of the impulse regime $t^*$ as a function of the protocol duration $\tau_q$ for a linear ramp of the coupling strength $g$ from the normal phase to the critical point $g_c$. The scaling exponent is found to be $\dot \Sigma_\mathrm{na}(t^*) \sim \tau_q^y$ with $y = -0.508 \pm 0.001$, in excellent agreement with the theoretical prediction $y = (\alpha-2-\gamma)/(\gamma+1)$ with $\alpha = 2$ and $\gamma =1$.  c) The nonadiabatic entropy production as a function of time for different protocol durations $\tau_q$ of a linear ramp of the coupling strength $g$ from the normal phase to the critical point $g_c$. We split the total nonadiabatic entropy production (black squared markers) into two contributions (see Eq.~\eqref{eq:na_quasi_impulse}), stemming from the quasiadiabatic (blue cross markers) and the impulse regime (red dot markers). Parameters: cavity frequency $\omega_c = 1$, atomic frequency $\omega_z = 1$, photon loss rate $\kappa = 0.2$, initial coupling strength $g_0 = g_c -  10^{-2}$, final coupling strength $g_f = g_c -  10^{-15}$.}
\label{fig:ODM_Sigmana_scaling}
\end{figure*}
{\bf Bosonic Gaussian DQPTs}.---We now highlight a remarkable feature of DQPTs in the mean-field, Ising-like universality class~\cite{Stanley_scaling_1999}. This includes paradigmatic fully connected open quantum systems exhibiting spontaneous weak $\mathbb{Z}_2$ symmetry breaking, such as the open Dicke~\cite{nagy_critical_2011}, Rabi~\cite{Hwang_dissipative_2018}, Kerr~\cite{zhang_driven-dissipative_2021}, and Lipkin–Meshkov–Glick~\cite{Ferreira_LMG_2019} models. These models, which are generally non-Gaussian, can be approximated near criticality as a bosonic Gaussian system with a single soft mode. In this case, we show in Appendix~\ref{app:KMB-scaling} that the QFI diverges with critical exponent $\alpha = 2$. Notably, this exponent is universal within bosonic Gaussian models, in the sense that it does not depend on the Liouvillian gap exponent or any other critical exponent that typically appears in scaling predictions, making it a distinctive feature of these systems.
This has a striking consequence for the integrated entropy production in both the adiabatic and impulse regimes.
To leading order, the total nonadiabatic entropy production is independent of the quench duration:
\begin{equation}
\Sigma_{\mathrm{na}}(0,\tau_q) \sim \tau_q^0.
\end{equation}
In other words, nonadiabatic effects always contribute to dissipation for such DQPTs, even for arbitrarily slow driving. This differs from the near-equilibrium physics of isolated, fully connected models, where the work dissipated by finite-speed driving across a QPT scales universally to zero as $\tau_q \to \infty$~\cite{Acevedo2014, hwang_quantum_2015}.
To validate these analytical scaling predictions, we consider the open quantum Dicke model (see Appendix~\ref{app:Dicke:model} for details), a paradigmatic example of a driven-dissipative system exhibiting a DQPT~\cite{dicke_coherence_1954, hepp_equilibrium_1973, wang_phase_1973, hioe_phase_1973, klinder_dynamical_2015, lang_critical_2016, paz_driven-dissipative_2022, dimer_proposed_2007, nagy_critical_2011, oztop_excitations_2012}. The model describes the collective interaction of a single-mode optical cavity field with a large ensemble of $N$ two-level atoms, subject to photon loss. In the thermodynamic limit, $N\to\infty$, it undergoes a transition from a normal to a superradiant phase. The latter is characterized by a macroscopic occupation of the cavity mode and collective atomic alignment, associated with $\mathbb{Z}_2$-symmetry breaking. 
In this limit, the system admits an effective Gaussian description~\cite{torre_keldysh_2013, kirton_introduction_2019, dimer_proposed_2007} with Hamiltonian (in the normal phase)
\begin{align}
\hat H(t) = \omega_c \hat a^\dagger \hat a + \omega_z \hat b^\dagger \hat b  + g(t) (\hat a + \hat a^\dagger)(\hat b + \hat b^\dagger). \label{eq:HPH_main_text}
\end{align}
Here, $\hat a$ and $\hat b$ are bosonic ladder operators describing the cavity and atomic degrees of freedom, respectively, $\omega_c$ is the cavity photon frequency, $\omega_z$ is the atomic energy splitting, and $g(t)$ denotes the time-dependent photon-atom coupling.
Cavity photon loss is modeled by the Lindblad equation
\begin{equation}
\frac{\mathrm{d}\hat{\rho}}{\mathrm{d}t} = -i[\hat H, \hat\rho] + \kappa \mathcal{D} [\hat a]\hat\rho ,
\end{equation}
where $\mathcal{D}[\hat L]\hat\rho = \hat L\hat\rho \hat L^\dagger - \frac{1}{2}\{\hat L^\dagger \hat L, \hat\rho\}$, $\kappa$ is the cavity decay rate, and we set $\hbar=1$.
Using this Gaussian model, we numerically confirm the predicted scaling of both the nonadiabatic entropy production rate and its integrated value in Fig.~\ref{fig:ODM_Sigmana_scaling}. In Appendix~\ref{app:Kerr:model}, we further substantiate our results via finite-size scaling in the non-Gaussian open Kerr model~\cite{zhang_driven-dissipative_2021}, demonstrating that our predictions are not merely an artefact of the Gaussian description that holds strictly in the thermodynamic limit.

{\bf Conclusion.}---We have developed a general framework for critical quantum dynamics far from equilibrium, extending the Kibble--Zurek paradigm to second-order DQPTs. By deriving scaling laws for the nonadiabatic entropy production, a measure of excess dissipation analogous to the dissipated work in closed systems~\cite{deffner_kibble-zurek_2017,fei_work_2020,naze_kibblezurek_2022,Ma2025, ShenQuantifyingMinimizing2025}, we have shown that the closing of the Liouvillian gap and the divergence of the Kubo-Mori-Bogoliubov QFI govern the universal behavior of irreversible thermodynamic signatures. The Fisher information metric has already been recognized as a powerful tool to identify critical points and map out phase diagrams even in the absence of a local order parameter~\cite{Ruppeiner_Riemannian_1995,Brody_Geometrical_1995,zanardi_ground_2006,venuti_quantum_2007,you_fidelity_2007,zanardi_information-theoretic_2007,banchi_quantum_2014,wang_quantum_2014,Marzolino_Fisher_2017}. 
By connecting the information geometry of quantum NESS to thermodynamic irreversibility and critical slowing down, our work establishes a new framework for characterizing dissipative universality classes beyond near-equilibrium and unitary settings.

Remarkably, for Gaussian bosonic systems with a single soft mode, the quadratic divergence of the QFI near criticality yields a total nonadiabatic entropy production that is, to leading order, independent of the driving speed. Our predictions are validated in the driven-dissipative Dicke model and confirmed by finite-size scaling in the open Kerr oscillator. Since the open Kerr, Rabi, and Lipkin-Meshkov-Glick models share the same universality class as the open Dicke model (see e.g. Ref.~\cite{Liu_universal_2025}), i.e.~that of the classical mean-field Ising model, they are expected to exhibit identical thermodynamic signatures near criticality, reflecting a common origin in the effective classical stochastic dynamics of a $\mathbb{Z}_2$-symmetric order parameter. 

Our results highlight the need for experimental probes of entropic currents in quantum devices. While stochastic entropy production has been measured in classical systems to verify fluctuation relations such as the Hatano–Sasa relation~\cite{Trepagnier2004, Mounier2012, Granger2015, Joubaud2008, Wang2002, hatano_steady-state_2001, esposito_three_2010}, only a few quantum experiments have accessed entropic currents directly~\cite{Koski2013, Hekking2013,pekola_towards_2015,brunelli_experimental_2018,  Rossi2020, Aguilar2022, aamirThermallyDrivenQuantum2025, Wadhia2025}. 
Looking ahead, our work also motivates time-resolved probing of other routinely measured currents, such as the photon flux in cavity-QED realizations of open Dicke or Kerr models, which we have shown to also exhibit Kibble–Zurek scaling~\cite{bettmann_inprep}. The cavity photon flux is directly related to the entropic flux~\cite{goes_quantum_2020} and has already been measured experimentally~\cite{brunelli_experimental_2018}. Because the nonadiabatic entropy production rate can be expressed in terms of both the rate of change of the von Neumann entropy and the excess entropy flux to the environment~\cite{lacerda_information_2025}, combined measurements of the Wigner function and cavity photon emisson may offer a feasible route to reconstructing the nonadiabatic entropy production in critical dissipative quantum dynamics.

\begin{acknowledgements}
We thank Klaus Mølmer, Anatoli Polkovnikov, Adolfo del Campo, Gabriel T. Landi, François Damanet, Eoin Carolan, Juan P. Garrahan, Fabrizio Minganti and Giacomo Guarnieri for fruitful discussions and feedback. LPB and JG acknowledge Research Ireland for support through the Frontiers for the Future project. MTM is supported by a Royal Society University Research Fellowship. JG is funded by a Research Ireland-Royal Society University Research Fellowship.
\end{acknowledgements}
\bibliographystyle{apsrev4-2}
\bibliography{references}

\begin{thebibliography}{100}%
\makeatletter
\providecommand \@ifxundefined [1]{%
 \@ifx{#1\undefined}
}%
\providecommand \@ifnum [1]{%
 \ifnum #1\expandafter \@firstoftwo
 \else \expandafter \@secondoftwo
 \fi
}%
\providecommand \@ifx [1]{%
 \ifx #1\expandafter \@firstoftwo
 \else \expandafter \@secondoftwo
 \fi
}%
\providecommand \natexlab [1]{#1}%
\providecommand \enquote  [1]{``#1''}%
\providecommand \bibnamefont  [1]{#1}%
\providecommand \bibfnamefont [1]{#1}%
\providecommand \citenamefont [1]{#1}%
\providecommand \href@noop [0]{\@secondoftwo}%
\providecommand \href [0]{\begingroup \@sanitize@url \@href}%
\providecommand \@href[1]{\@@startlink{#1}\@@href}%
\providecommand \@@href[1]{\endgroup#1\@@endlink}%
\providecommand \@sanitize@url [0]{\catcode `\\12\catcode `\$12\catcode `\&12\catcode `\#12\catcode `\^12\catcode `\_12\catcode `\%12\relax}%
\providecommand \@@startlink[1]{}%
\providecommand \@@endlink[0]{}%
\providecommand \url  [0]{\begingroup\@sanitize@url \@url }%
\providecommand \@url [1]{\endgroup\@href {#1}{\urlprefix }}%
\providecommand \urlprefix  [0]{URL }%
\providecommand \Eprint [0]{\href }%
\providecommand \doibase [0]{https://doi.org/}%
\providecommand \selectlanguage [0]{\@gobble}%
\providecommand \bibinfo  [0]{\@secondoftwo}%
\providecommand \bibfield  [0]{\@secondoftwo}%
\providecommand \translation [1]{[#1]}%
\providecommand \BibitemOpen [0]{}%
\providecommand \bibitemStop [0]{}%
\providecommand \bibitemNoStop [0]{.\EOS\space}%
\providecommand \EOS [0]{\spacefactor3000\relax}%
\providecommand \BibitemShut  [1]{\csname bibitem#1\endcsname}%
\let\auto@bib@innerbib\@empty
\bibitem [{\citenamefont {Kibble}(1976)}]{Kibble1976}%
  \BibitemOpen
  \bibfield  {author} {\bibinfo {author} {\bibfnamefont {T.~W.~B.}\ \bibnamefont {Kibble}},\ }\href {https://doi.org/10.1088/0305-4470/9/8/029} {\bibfield  {journal} {\bibinfo  {journal} {J. Phys. A: Math. Gen.}\ }\textbf {\bibinfo {volume} {9}},\ \bibinfo {pages} {1387} (\bibinfo {year} {1976})}\BibitemShut {NoStop}%
\bibitem [{\citenamefont {Zurek}(1985)}]{Zurek1985}%
  \BibitemOpen
  \bibfield  {author} {\bibinfo {author} {\bibfnamefont {W.~H.}\ \bibnamefont {Zurek}},\ }\href {https://doi.org/10.1038/317505a0} {\bibfield  {journal} {\bibinfo  {journal} {Nature}\ }\textbf {\bibinfo {volume} {317}},\ \bibinfo {pages} {505} (\bibinfo {year} {1985})}\BibitemShut {NoStop}%
\bibitem [{\citenamefont {Polkovnikov}\ \emph {et~al.}(2011)\citenamefont {Polkovnikov}, \citenamefont {Sengupta}, \citenamefont {Silva},\ and\ \citenamefont {Vengalattore}}]{polkovnikov_rmp_2011}%
  \BibitemOpen
  \bibfield  {author} {\bibinfo {author} {\bibfnamefont {A.}~\bibnamefont {Polkovnikov}}, \bibinfo {author} {\bibfnamefont {K.}~\bibnamefont {Sengupta}}, \bibinfo {author} {\bibfnamefont {A.}~\bibnamefont {Silva}},\ and\ \bibinfo {author} {\bibfnamefont {M.}~\bibnamefont {Vengalattore}},\ }\href {https://doi.org/10.1103/RevModPhys.83.863} {\bibfield  {journal} {\bibinfo  {journal} {Rev. Mod. Phys.}\ }\textbf {\bibinfo {volume} {83}},\ \bibinfo {pages} {863} (\bibinfo {year} {2011})}\BibitemShut {NoStop}%
\bibitem [{\citenamefont {del Campo}\ and\ \citenamefont {Zurek}(2014)}]{delCampo2014}%
  \BibitemOpen
  \bibfield  {author} {\bibinfo {author} {\bibfnamefont {A.}~\bibnamefont {del Campo}}\ and\ \bibinfo {author} {\bibfnamefont {W.~H.}\ \bibnamefont {Zurek}},\ }\href {https://doi.org/10.1142/S0217751X1430018X} {\bibfield  {journal} {\bibinfo  {journal} {Int. J. Mod. Phys. A}\ }\textbf {\bibinfo {volume} {29}},\ \bibinfo {pages} {1430018} (\bibinfo {year} {2014})}\BibitemShut {NoStop}%
\bibitem [{\citenamefont {Dziarmaga}(2010)}]{Dziarmaga2010}%
  \BibitemOpen
  \bibfield  {author} {\bibinfo {author} {\bibfnamefont {J.}~\bibnamefont {Dziarmaga}},\ }\href {https://doi.org/10.1080/00018732.2010.514702} {\bibfield  {journal} {\bibinfo  {journal} {Adv. Phys.}\ }\textbf {\bibinfo {volume} {59}},\ \bibinfo {pages} {1063} (\bibinfo {year} {2010})}\BibitemShut {NoStop}%
\bibitem [{\citenamefont {Biroli}\ \emph {et~al.}(2010)\citenamefont {Biroli}, \citenamefont {Cugliandolo},\ and\ \citenamefont {Sicilia}}]{biroli_kibble-zurek_2010}%
  \BibitemOpen
  \bibfield  {author} {\bibinfo {author} {\bibfnamefont {G.}~\bibnamefont {Biroli}}, \bibinfo {author} {\bibfnamefont {L.~F.}\ \bibnamefont {Cugliandolo}},\ and\ \bibinfo {author} {\bibfnamefont {A.}~\bibnamefont {Sicilia}},\ }\href {https://doi.org/10.1103/PhysRevE.81.050101} {\bibfield  {journal} {\bibinfo  {journal} {Physical Review E}\ }\textbf {\bibinfo {volume} {81}},\ \bibinfo {pages} {050101} (\bibinfo {year} {2010})}\BibitemShut {NoStop}%
\bibitem [{\citenamefont {Sachdev}(2011)}]{Sachdev2011}%
  \BibitemOpen
  \bibfield  {author} {\bibinfo {author} {\bibfnamefont {S.}~\bibnamefont {Sachdev}},\ }\href@noop {} {\emph {\bibinfo {title} {Quantum Phase Transitions}}},\ \bibinfo {edition} {2nd}\ ed.\ (\bibinfo  {publisher} {Cambridge University Press},\ \bibinfo {address} {Cambridge, England},\ \bibinfo {year} {2011})\BibitemShut {NoStop}%
\bibitem [{\citenamefont {Zurek}\ \emph {et~al.}(2005)\citenamefont {Zurek}, \citenamefont {Dorner},\ and\ \citenamefont {Zoller}}]{zurek_dynamics_2005}%
  \BibitemOpen
  \bibfield  {author} {\bibinfo {author} {\bibfnamefont {W.~H.}\ \bibnamefont {Zurek}}, \bibinfo {author} {\bibfnamefont {U.}~\bibnamefont {Dorner}},\ and\ \bibinfo {author} {\bibfnamefont {P.}~\bibnamefont {Zoller}},\ }\href {https://doi.org/10.1103/PhysRevLett.95.105701} {\bibfield  {journal} {\bibinfo  {journal} {Physical Review Letters}\ }\textbf {\bibinfo {volume} {95}},\ \bibinfo {pages} {105701} (\bibinfo {year} {2005})}\BibitemShut {NoStop}%
\bibitem [{\citenamefont {Polkovnikov}(2005)}]{Polkovnikov2005}%
  \BibitemOpen
  \bibfield  {author} {\bibinfo {author} {\bibfnamefont {A.}~\bibnamefont {Polkovnikov}},\ }\href {https://doi.org/10.1103/PhysRevB.72.161201} {\bibfield  {journal} {\bibinfo  {journal} {Phys. Rev. B}\ }\textbf {\bibinfo {volume} {72}},\ \bibinfo {pages} {161201} (\bibinfo {year} {2005})}\BibitemShut {NoStop}%
\bibitem [{\citenamefont {Deffner}(2017)}]{deffner_kibble-zurek_2017}%
  \BibitemOpen
  \bibfield  {author} {\bibinfo {author} {\bibfnamefont {S.}~\bibnamefont {Deffner}},\ }\href {https://doi.org/10.1103/PhysRevE.96.052125} {\bibfield  {journal} {\bibinfo  {journal} {Physical Review E}\ }\textbf {\bibinfo {volume} {96}},\ \bibinfo {pages} {052125} (\bibinfo {year} {2017})}\BibitemShut {NoStop}%
\bibitem [{\citenamefont {Fei}\ \emph {et~al.}(2020)\citenamefont {Fei}, \citenamefont {Freitas}, \citenamefont {Cavina}, \citenamefont {Quan},\ and\ \citenamefont {Esposito}}]{fei_work_2020}%
  \BibitemOpen
  \bibfield  {author} {\bibinfo {author} {\bibfnamefont {Z.}~\bibnamefont {Fei}}, \bibinfo {author} {\bibfnamefont {N.}~\bibnamefont {Freitas}}, \bibinfo {author} {\bibfnamefont {V.}~\bibnamefont {Cavina}}, \bibinfo {author} {\bibfnamefont {H.}~\bibnamefont {Quan}},\ and\ \bibinfo {author} {\bibfnamefont {M.}~\bibnamefont {Esposito}},\ }\href {https://doi.org/10.1103/PhysRevLett.124.170603} {\bibfield  {journal} {\bibinfo  {journal} {Physical Review Letters}\ }\textbf {\bibinfo {volume} {124}},\ \bibinfo {pages} {170603} (\bibinfo {year} {2020})}\BibitemShut {NoStop}%
\bibitem [{\citenamefont {Nazé}\ \emph {et~al.}(2022)\citenamefont {Nazé}, \citenamefont {Bonança},\ and\ \citenamefont {Deffner}}]{naze_kibblezurek_2022}%
  \BibitemOpen
  \bibfield  {author} {\bibinfo {author} {\bibfnamefont {P.}~\bibnamefont {Nazé}}, \bibinfo {author} {\bibfnamefont {M.~V.~S.}\ \bibnamefont {Bonança}},\ and\ \bibinfo {author} {\bibfnamefont {S.}~\bibnamefont {Deffner}},\ }\href {https://doi.org/10.3390/e24050666} {\bibfield  {journal} {\bibinfo  {journal} {Entropy}\ }\textbf {\bibinfo {volume} {24}},\ \bibinfo {pages} {666} (\bibinfo {year} {2022})}\BibitemShut {NoStop}%
\bibitem [{\citenamefont {Ma}\ \emph {et~al.}(2025)\citenamefont {Ma}, \citenamefont {Mitchell},\ and\ \citenamefont {Sela}}]{Ma2025}%
  \BibitemOpen
  \bibfield  {author} {\bibinfo {author} {\bibfnamefont {Z.}~\bibnamefont {Ma}}, \bibinfo {author} {\bibfnamefont {A.~K.}\ \bibnamefont {Mitchell}},\ and\ \bibinfo {author} {\bibfnamefont {E.}~\bibnamefont {Sela}},\ }\href {https://doi.org/10.1103/vn83-mt2v} {\bibfield  {journal} {\bibinfo  {journal} {Phys. Rev. Lett.}\ }\textbf {\bibinfo {volume} {135}},\ \bibinfo {pages} {130402} (\bibinfo {year} {2025})}\BibitemShut {NoStop}%
\bibitem [{\citenamefont {Shen}\ \emph {et~al.}(2025)\citenamefont {Shen}, \citenamefont {Jiang}, \citenamefont {Huang}, \citenamefont {Cleary}, \citenamefont {Jiang}, \citenamefont {Rotskoff},\ and\ \citenamefont {Lindenberg}}]{ShenQuantifyingMinimizing2025}%
  \BibitemOpen
  \bibfield  {author} {\bibinfo {author} {\bibfnamefont {Y.}~\bibnamefont {Shen}}, \bibinfo {author} {\bibfnamefont {Z.}~\bibnamefont {Jiang}}, \bibinfo {author} {\bibfnamefont {Y.}~\bibnamefont {Huang}}, \bibinfo {author} {\bibfnamefont {B.~M.}\ \bibnamefont {Cleary}}, \bibinfo {author} {\bibfnamefont {Y.}~\bibnamefont {Jiang}}, \bibinfo {author} {\bibfnamefont {G.~M.}\ \bibnamefont {Rotskoff}},\ and\ \bibinfo {author} {\bibfnamefont {A.~M.}\ \bibnamefont {Lindenberg}},\ }\href {https://doi.org/10.48550/arXiv.2511.12814} {\bibinfo {title} {Quantifying and minimizing dissipation in a non-equilibrium phase transition}} (\bibinfo {year} {2025}),\ \Eprint {https://arxiv.org/abs/2511.12814} {arXiv:2511.12814 [cond-mat]} \BibitemShut {NoStop}%
\bibitem [{\citenamefont {Sieberer}\ \emph {et~al.}(2025)\citenamefont {Sieberer}, \citenamefont {Buchhold}, \citenamefont {Marino},\ and\ \citenamefont {Diehl}}]{Sieberer_universality_2025}%
  \BibitemOpen
  \bibfield  {author} {\bibinfo {author} {\bibfnamefont {L.~M.}\ \bibnamefont {Sieberer}}, \bibinfo {author} {\bibfnamefont {M.}~\bibnamefont {Buchhold}}, \bibinfo {author} {\bibfnamefont {J.}~\bibnamefont {Marino}},\ and\ \bibinfo {author} {\bibfnamefont {S.}~\bibnamefont {Diehl}},\ }\href {https://doi.org/10.1103/RevModPhys.97.025004} {\bibfield  {journal} {\bibinfo  {journal} {Rev. Mod. Phys.}\ }\textbf {\bibinfo {volume} {97}},\ \bibinfo {pages} {025004} (\bibinfo {year} {2025})}\BibitemShut {NoStop}%
\bibitem [{\citenamefont {Baumann}\ \emph {et~al.}(2010)\citenamefont {Baumann}, \citenamefont {Guerlin}, \citenamefont {Brennecke},\ and\ \citenamefont {Esslinger}}]{Baumann2010}%
  \BibitemOpen
  \bibfield  {author} {\bibinfo {author} {\bibfnamefont {K.}~\bibnamefont {Baumann}}, \bibinfo {author} {\bibfnamefont {C.}~\bibnamefont {Guerlin}}, \bibinfo {author} {\bibfnamefont {F.}~\bibnamefont {Brennecke}},\ and\ \bibinfo {author} {\bibfnamefont {T.}~\bibnamefont {Esslinger}},\ }\href {https://doi.org/10.1038/nature09009} {\bibfield  {journal} {\bibinfo  {journal} {Nature}\ }\textbf {\bibinfo {volume} {464}},\ \bibinfo {pages} {1301} (\bibinfo {year} {2010})}\BibitemShut {NoStop}%
\bibitem [{\citenamefont {Ferri}\ \emph {et~al.}(2021)\citenamefont {Ferri}, \citenamefont {{Rosa-Medina}}, \citenamefont {Finger}, \citenamefont {Dogra}, \citenamefont {Soriente}, \citenamefont {Zilberberg}, \citenamefont {Donner},\ and\ \citenamefont {Esslinger}}]{Ferri2021}%
  \BibitemOpen
  \bibfield  {author} {\bibinfo {author} {\bibfnamefont {F.}~\bibnamefont {Ferri}}, \bibinfo {author} {\bibfnamefont {R.}~\bibnamefont {{Rosa-Medina}}}, \bibinfo {author} {\bibfnamefont {F.}~\bibnamefont {Finger}}, \bibinfo {author} {\bibfnamefont {N.}~\bibnamefont {Dogra}}, \bibinfo {author} {\bibfnamefont {M.}~\bibnamefont {Soriente}}, \bibinfo {author} {\bibfnamefont {O.}~\bibnamefont {Zilberberg}}, \bibinfo {author} {\bibfnamefont {T.}~\bibnamefont {Donner}},\ and\ \bibinfo {author} {\bibfnamefont {T.}~\bibnamefont {Esslinger}},\ }\href {https://doi.org/10.1103/PhysRevX.11.041046} {\bibfield  {journal} {\bibinfo  {journal} {Physical Review X}\ }\textbf {\bibinfo {volume} {11}},\ \bibinfo {pages} {041046} (\bibinfo {year} {2021})}\BibitemShut {NoStop}%
\bibitem [{\citenamefont {Fitzpatrick}\ \emph {et~al.}(2017)\citenamefont {Fitzpatrick}, \citenamefont {Sundaresan}, \citenamefont {Li}, \citenamefont {Koch},\ and\ \citenamefont {Houck}}]{fitzpatrick_observation_2017}%
  \BibitemOpen
  \bibfield  {author} {\bibinfo {author} {\bibfnamefont {M.}~\bibnamefont {Fitzpatrick}}, \bibinfo {author} {\bibfnamefont {N.~M.}\ \bibnamefont {Sundaresan}}, \bibinfo {author} {\bibfnamefont {A.~C.}\ \bibnamefont {Li}}, \bibinfo {author} {\bibfnamefont {J.}~\bibnamefont {Koch}},\ and\ \bibinfo {author} {\bibfnamefont {A.~A.}\ \bibnamefont {Houck}},\ }\href {https://doi.org/10.1103/PhysRevX.7.011016} {\bibfield  {journal} {\bibinfo  {journal} {Physical Review X}\ }\textbf {\bibinfo {volume} {7}},\ \bibinfo {pages} {011016} (\bibinfo {year} {2017})}\BibitemShut {NoStop}%
\bibitem [{\citenamefont {Rodriguez}\ \emph {et~al.}(2017)\citenamefont {Rodriguez}, \citenamefont {Casteels}, \citenamefont {Storme}, \citenamefont {Carlon~Zambon}, \citenamefont {Sagnes}, \citenamefont {Le~Gratiet}, \citenamefont {Galopin}, \citenamefont {Lemaître}, \citenamefont {Amo}, \citenamefont {Ciuti},\ and\ \citenamefont {Bloch}}]{rodriguez_probing_2017}%
  \BibitemOpen
  \bibfield  {author} {\bibinfo {author} {\bibfnamefont {S.}~\bibnamefont {Rodriguez}}, \bibinfo {author} {\bibfnamefont {W.}~\bibnamefont {Casteels}}, \bibinfo {author} {\bibfnamefont {F.}~\bibnamefont {Storme}}, \bibinfo {author} {\bibfnamefont {N.}~\bibnamefont {Carlon~Zambon}}, \bibinfo {author} {\bibfnamefont {I.}~\bibnamefont {Sagnes}}, \bibinfo {author} {\bibfnamefont {L.}~\bibnamefont {Le~Gratiet}}, \bibinfo {author} {\bibfnamefont {E.}~\bibnamefont {Galopin}}, \bibinfo {author} {\bibfnamefont {A.}~\bibnamefont {Lemaître}}, \bibinfo {author} {\bibfnamefont {A.}~\bibnamefont {Amo}}, \bibinfo {author} {\bibfnamefont {C.}~\bibnamefont {Ciuti}},\ and\ \bibinfo {author} {\bibfnamefont {J.}~\bibnamefont {Bloch}},\ }\href {https://doi.org/10.1103/PhysRevLett.118.247402} {\bibfield  {journal} {\bibinfo  {journal} {Physical Review Letters}\ }\textbf {\bibinfo {volume} {118}},\ \bibinfo {pages} {247402} (\bibinfo {year} {2017})}\BibitemShut {NoStop}%
\bibitem [{\citenamefont {Fink}\ \emph {et~al.}(2018)\citenamefont {Fink}, \citenamefont {Schade}, \citenamefont {Höfling}, \citenamefont {Schneider},\ and\ \citenamefont {Imamoglu}}]{fink_signatures_2018}%
  \BibitemOpen
  \bibfield  {author} {\bibinfo {author} {\bibfnamefont {T.}~\bibnamefont {Fink}}, \bibinfo {author} {\bibfnamefont {A.}~\bibnamefont {Schade}}, \bibinfo {author} {\bibfnamefont {S.}~\bibnamefont {Höfling}}, \bibinfo {author} {\bibfnamefont {C.}~\bibnamefont {Schneider}},\ and\ \bibinfo {author} {\bibfnamefont {A.}~\bibnamefont {Imamoglu}},\ }\href {https://doi.org/10.1038/s41567-017-0020-9} {\bibfield  {journal} {\bibinfo  {journal} {Nature Physics}\ }\textbf {\bibinfo {volume} {14}},\ \bibinfo {pages} {365} (\bibinfo {year} {2018})}\BibitemShut {NoStop}%
\bibitem [{\citenamefont {Beaulieu}\ \emph {et~al.}(2025)\citenamefont {Beaulieu}, \citenamefont {Minganti}, \citenamefont {Frasca}, \citenamefont {Savona}, \citenamefont {Felicetti}, \citenamefont {Di~Candia},\ and\ \citenamefont {Scarlino}}]{beaulieu_observation_2025}%
  \BibitemOpen
  \bibfield  {author} {\bibinfo {author} {\bibfnamefont {G.}~\bibnamefont {Beaulieu}}, \bibinfo {author} {\bibfnamefont {F.}~\bibnamefont {Minganti}}, \bibinfo {author} {\bibfnamefont {S.}~\bibnamefont {Frasca}}, \bibinfo {author} {\bibfnamefont {V.}~\bibnamefont {Savona}}, \bibinfo {author} {\bibfnamefont {S.}~\bibnamefont {Felicetti}}, \bibinfo {author} {\bibfnamefont {R.}~\bibnamefont {Di~Candia}},\ and\ \bibinfo {author} {\bibfnamefont {P.}~\bibnamefont {Scarlino}},\ }\href {https://doi.org/10.1038/s41467-025-56830-w} {\bibfield  {journal} {\bibinfo  {journal} {Nature Communications}\ }\textbf {\bibinfo {volume} {16}},\ \bibinfo {pages} {1954} (\bibinfo {year} {2025})}\BibitemShut {NoStop}%
\bibitem [{\citenamefont {Brennecke}\ \emph {et~al.}(2013)\citenamefont {Brennecke}, \citenamefont {Mottl}, \citenamefont {Baumann}, \citenamefont {Landig}, \citenamefont {Donner},\ and\ \citenamefont {Esslinger}}]{Brennecke2013}%
  \BibitemOpen
  \bibfield  {author} {\bibinfo {author} {\bibfnamefont {F.}~\bibnamefont {Brennecke}}, \bibinfo {author} {\bibfnamefont {R.}~\bibnamefont {Mottl}}, \bibinfo {author} {\bibfnamefont {K.}~\bibnamefont {Baumann}}, \bibinfo {author} {\bibfnamefont {R.}~\bibnamefont {Landig}}, \bibinfo {author} {\bibfnamefont {T.}~\bibnamefont {Donner}},\ and\ \bibinfo {author} {\bibfnamefont {T.}~\bibnamefont {Esslinger}},\ }\href {https://doi.org/10.1073/pnas.1306993110} {\bibfield  {journal} {\bibinfo  {journal} {Proceedings of the National Academy of Sciences}\ }\textbf {\bibinfo {volume} {110}},\ \bibinfo {pages} {11763} (\bibinfo {year} {2013})}\BibitemShut {NoStop}%
\bibitem [{\citenamefont {Brunelli}\ \emph {et~al.}(2018)\citenamefont {Brunelli}, \citenamefont {Fusco}, \citenamefont {Landig}, \citenamefont {Wieczorek}, \citenamefont {Hoelscher-Obermaier}, \citenamefont {Landi}, \citenamefont {Semiao}, \citenamefont {Ferraro}, \citenamefont {Kiesel}, \citenamefont {Donner}, \citenamefont {Chiara},\ and\ \citenamefont {Paternostro}}]{brunelli_experimental_2018}%
  \BibitemOpen
  \bibfield  {author} {\bibinfo {author} {\bibfnamefont {M.}~\bibnamefont {Brunelli}}, \bibinfo {author} {\bibfnamefont {L.}~\bibnamefont {Fusco}}, \bibinfo {author} {\bibfnamefont {R.}~\bibnamefont {Landig}}, \bibinfo {author} {\bibfnamefont {W.}~\bibnamefont {Wieczorek}}, \bibinfo {author} {\bibfnamefont {J.}~\bibnamefont {Hoelscher-Obermaier}}, \bibinfo {author} {\bibfnamefont {G.}~\bibnamefont {Landi}}, \bibinfo {author} {\bibfnamefont {F.~L.}\ \bibnamefont {Semiao}}, \bibinfo {author} {\bibfnamefont {A.}~\bibnamefont {Ferraro}}, \bibinfo {author} {\bibfnamefont {N.}~\bibnamefont {Kiesel}}, \bibinfo {author} {\bibfnamefont {T.}~\bibnamefont {Donner}}, \bibinfo {author} {\bibfnamefont {G.~D.}\ \bibnamefont {Chiara}},\ and\ \bibinfo {author} {\bibfnamefont {M.}~\bibnamefont {Paternostro}},\ }\href {https://doi.org/10.1103/PhysRevLett.121.160604} {\bibfield  {journal} {\bibinfo  {journal} {Physical Review Letters}\ }\textbf {\bibinfo {volume} {121}},\ \bibinfo {pages} {160604} (\bibinfo {year}
  {2018})}\BibitemShut {NoStop}%
\bibitem [{\citenamefont {Ducci}\ \emph {et~al.}(1999)\citenamefont {Ducci}, \citenamefont {Ramazza}, \citenamefont {González-Viñas},\ and\ \citenamefont {Arecchi}}]{ducci_order_1999}%
  \BibitemOpen
  \bibfield  {author} {\bibinfo {author} {\bibfnamefont {S.}~\bibnamefont {Ducci}}, \bibinfo {author} {\bibfnamefont {P.~L.}\ \bibnamefont {Ramazza}}, \bibinfo {author} {\bibfnamefont {W.}~\bibnamefont {González-Viñas}},\ and\ \bibinfo {author} {\bibfnamefont {F.~T.}\ \bibnamefont {Arecchi}},\ }\href {https://doi.org/10.1103/PhysRevLett.83.5210} {\bibfield  {journal} {\bibinfo  {journal} {Physical Review Letters}\ }\textbf {\bibinfo {volume} {83}},\ \bibinfo {pages} {5210} (\bibinfo {year} {1999})}\BibitemShut {NoStop}%
\bibitem [{\citenamefont {Casado}\ \emph {et~al.}(2007)\citenamefont {Casado}, \citenamefont {González-Viñas}, \citenamefont {Boccaletti}, \citenamefont {Ramazza},\ and\ \citenamefont {Mancini}}]{casado_birth_2007}%
  \BibitemOpen
  \bibfield  {author} {\bibinfo {author} {\bibfnamefont {S.}~\bibnamefont {Casado}}, \bibinfo {author} {\bibfnamefont {W.}~\bibnamefont {González-Viñas}}, \bibinfo {author} {\bibfnamefont {S.}~\bibnamefont {Boccaletti}}, \bibinfo {author} {\bibfnamefont {P.~L.}\ \bibnamefont {Ramazza}},\ and\ \bibinfo {author} {\bibfnamefont {H.}~\bibnamefont {Mancini}},\ }\href {https://doi.org/10.1140/epjst/e2007-00171-2} {\bibfield  {journal} {\bibinfo  {journal} {The European Physical Journal Special Topics}\ }\textbf {\bibinfo {volume} {146}},\ \bibinfo {pages} {87} (\bibinfo {year} {2007})}\BibitemShut {NoStop}%
\bibitem [{\citenamefont {Casado}\ \emph {et~al.}(2006)\citenamefont {Casado}, \citenamefont {González-Viñas},\ and\ \citenamefont {Mancini}}]{casado_testing_2006}%
  \BibitemOpen
  \bibfield  {author} {\bibinfo {author} {\bibfnamefont {S.}~\bibnamefont {Casado}}, \bibinfo {author} {\bibfnamefont {W.}~\bibnamefont {González-Viñas}},\ and\ \bibinfo {author} {\bibfnamefont {H.}~\bibnamefont {Mancini}},\ }\href {https://doi.org/10.1103/PhysRevE.74.047101} {\bibfield  {journal} {\bibinfo  {journal} {Physical Review E}\ }\textbf {\bibinfo {volume} {74}},\ \bibinfo {pages} {047101} (\bibinfo {year} {2006})}\BibitemShut {NoStop}%
\bibitem [{\citenamefont {Casado}\ \emph {et~al.}(2001)\citenamefont {Casado}, \citenamefont {González-Viñas}, \citenamefont {Mancini},\ and\ \citenamefont {Boccaletti}}]{casado_topological_2001}%
  \BibitemOpen
  \bibfield  {author} {\bibinfo {author} {\bibfnamefont {S.}~\bibnamefont {Casado}}, \bibinfo {author} {\bibfnamefont {W.}~\bibnamefont {González-Viñas}}, \bibinfo {author} {\bibfnamefont {H.}~\bibnamefont {Mancini}},\ and\ \bibinfo {author} {\bibfnamefont {S.}~\bibnamefont {Boccaletti}},\ }\href {https://doi.org/10.1103/PhysRevE.63.057301} {\bibfield  {journal} {\bibinfo  {journal} {Physical Review E}\ }\textbf {\bibinfo {volume} {63}},\ \bibinfo {pages} {057301} (\bibinfo {year} {2001})}\BibitemShut {NoStop}%
\bibitem [{\citenamefont {Klinder}\ \emph {et~al.}(2015)\citenamefont {Klinder}, \citenamefont {Keßler}, \citenamefont {Wolke}, \citenamefont {Mathey},\ and\ \citenamefont {Hemmerich}}]{klinder_dynamical_2015}%
  \BibitemOpen
  \bibfield  {author} {\bibinfo {author} {\bibfnamefont {J.}~\bibnamefont {Klinder}}, \bibinfo {author} {\bibfnamefont {H.}~\bibnamefont {Keßler}}, \bibinfo {author} {\bibfnamefont {M.}~\bibnamefont {Wolke}}, \bibinfo {author} {\bibfnamefont {L.}~\bibnamefont {Mathey}},\ and\ \bibinfo {author} {\bibfnamefont {A.}~\bibnamefont {Hemmerich}},\ }\href {https://doi.org/10.1073/pnas.1417132112} {\bibfield  {journal} {\bibinfo  {journal} {Proceedings of the National Academy of Sciences}\ }\textbf {\bibinfo {volume} {112}},\ \bibinfo {pages} {3290} (\bibinfo {year} {2015})}\BibitemShut {NoStop}%
\bibitem [{\citenamefont {Zamora}\ \emph {et~al.}(2020)\citenamefont {Zamora}, \citenamefont {Dagvadorj}, \citenamefont {Comaron}, \citenamefont {Carusotto}, \citenamefont {Proukakis},\ and\ \citenamefont {Szymańska}}]{zamora_kibble-zurek_2020}%
  \BibitemOpen
  \bibfield  {author} {\bibinfo {author} {\bibfnamefont {A.}~\bibnamefont {Zamora}}, \bibinfo {author} {\bibfnamefont {G.}~\bibnamefont {Dagvadorj}}, \bibinfo {author} {\bibfnamefont {P.}~\bibnamefont {Comaron}}, \bibinfo {author} {\bibfnamefont {I.}~\bibnamefont {Carusotto}}, \bibinfo {author} {\bibfnamefont {N.}~\bibnamefont {Proukakis}},\ and\ \bibinfo {author} {\bibfnamefont {M.}~\bibnamefont {Szymańska}},\ }\href {https://doi.org/10.1103/PhysRevLett.125.095301} {\bibfield  {journal} {\bibinfo  {journal} {Physical Review Letters}\ }\textbf {\bibinfo {volume} {125}},\ \bibinfo {pages} {095301} (\bibinfo {year} {2020})}\BibitemShut {NoStop}%
\bibitem [{\citenamefont {Hedvall}\ and\ \citenamefont {Larson}(2017)}]{hedvall_dynamics_2017}%
  \BibitemOpen
  \bibfield  {author} {\bibinfo {author} {\bibfnamefont {P.}~\bibnamefont {Hedvall}}\ and\ \bibinfo {author} {\bibfnamefont {J.}~\bibnamefont {Larson}},\ }\href {https://doi.org/10.48550/arXiv.1712.01560} {\bibinfo {title} {Dynamics of non-equilibrium steady state quantum phase transitions}} (\bibinfo {year} {2017})\BibitemShut {NoStop}%
\bibitem [{\citenamefont {Lacerda}\ \emph {et~al.}(2025)\citenamefont {Lacerda}, \citenamefont {Bettmann},\ and\ \citenamefont {Goold}}]{lacerda_information_2025}%
  \BibitemOpen
  \bibfield  {author} {\bibinfo {author} {\bibfnamefont {A.~M.}\ \bibnamefont {Lacerda}}, \bibinfo {author} {\bibfnamefont {L.~P.}\ \bibnamefont {Bettmann}},\ and\ \bibinfo {author} {\bibfnamefont {J.}~\bibnamefont {Goold}},\ }\href {https://doi.org/10.1103/9f6l-d766} {\bibfield  {journal} {\bibinfo  {journal} {Phys. Rev. E}\ }\textbf {\bibinfo {volume} {112}},\ \bibinfo {pages} {L022101} (\bibinfo {year} {2025})}\BibitemShut {NoStop}%
\bibitem [{\citenamefont {Verstraelen}\ and\ \citenamefont {Wouters}(2020)}]{verstraelen_classical_2020}%
  \BibitemOpen
  \bibfield  {author} {\bibinfo {author} {\bibfnamefont {W.}~\bibnamefont {Verstraelen}}\ and\ \bibinfo {author} {\bibfnamefont {M.}~\bibnamefont {Wouters}},\ }\href {https://doi.org/10.1103/PhysRevA.101.043826} {\bibfield  {journal} {\bibinfo  {journal} {Physical Review A}\ }\textbf {\bibinfo {volume} {101}},\ \bibinfo {pages} {043826} (\bibinfo {year} {2020})}\BibitemShut {NoStop}%
\bibitem [{\citenamefont {Zhang}\ and\ \citenamefont {Baranger}(2021)}]{zhang_driven-dissipative_2021}%
  \BibitemOpen
  \bibfield  {author} {\bibinfo {author} {\bibfnamefont {X.~H.~H.}\ \bibnamefont {Zhang}}\ and\ \bibinfo {author} {\bibfnamefont {H.~U.}\ \bibnamefont {Baranger}},\ }\href {https://doi.org/10.1103/PhysRevA.103.033711} {\bibfield  {journal} {\bibinfo  {journal} {Physical Review A}\ }\textbf {\bibinfo {volume} {103}},\ \bibinfo {pages} {033711} (\bibinfo {year} {2021})}\BibitemShut {NoStop}%
\bibitem [{\citenamefont {Kessler}\ \emph {et~al.}(2012)\citenamefont {Kessler}, \citenamefont {Giedke}, \citenamefont {Imamoglu}, \citenamefont {Yelin}, \citenamefont {Lukin},\ and\ \citenamefont {Cirac}}]{kessler_dissipative_2012}%
  \BibitemOpen
  \bibfield  {author} {\bibinfo {author} {\bibfnamefont {E.~M.}\ \bibnamefont {Kessler}}, \bibinfo {author} {\bibfnamefont {G.}~\bibnamefont {Giedke}}, \bibinfo {author} {\bibfnamefont {A.}~\bibnamefont {Imamoglu}}, \bibinfo {author} {\bibfnamefont {S.~F.}\ \bibnamefont {Yelin}}, \bibinfo {author} {\bibfnamefont {M.~D.}\ \bibnamefont {Lukin}},\ and\ \bibinfo {author} {\bibfnamefont {J.~I.}\ \bibnamefont {Cirac}},\ }\href {https://doi.org/10.1103/PhysRevA.86.012116} {\bibfield  {journal} {\bibinfo  {journal} {Physical Review A}\ }\textbf {\bibinfo {volume} {86}},\ \bibinfo {pages} {012116} (\bibinfo {year} {2012})}\BibitemShut {NoStop}%
\bibitem [{\citenamefont {Torre}\ \emph {et~al.}(2013)\citenamefont {Torre}, \citenamefont {Diehl}, \citenamefont {Lukin}, \citenamefont {Sachdev},\ and\ \citenamefont {Strack}}]{torre_keldysh_2013}%
  \BibitemOpen
  \bibfield  {author} {\bibinfo {author} {\bibfnamefont {E.~G.~D.}\ \bibnamefont {Torre}}, \bibinfo {author} {\bibfnamefont {S.}~\bibnamefont {Diehl}}, \bibinfo {author} {\bibfnamefont {M.~D.}\ \bibnamefont {Lukin}}, \bibinfo {author} {\bibfnamefont {S.}~\bibnamefont {Sachdev}},\ and\ \bibinfo {author} {\bibfnamefont {P.}~\bibnamefont {Strack}},\ }\href {https://doi.org/10.1103/PhysRevA.87.023831} {\bibfield  {journal} {\bibinfo  {journal} {Physical Review A}\ }\textbf {\bibinfo {volume} {87}},\ \bibinfo {pages} {023831} (\bibinfo {year} {2013})}\BibitemShut {NoStop}%
\bibitem [{\citenamefont {Hatano}\ and\ \citenamefont {Sasa}(2001{\natexlab{a}})}]{hatano_steadystate_2001}%
  \BibitemOpen
  \bibfield  {author} {\bibinfo {author} {\bibfnamefont {T.}~\bibnamefont {Hatano}}\ and\ \bibinfo {author} {\bibfnamefont {S.-i.}\ \bibnamefont {Sasa}},\ }\href {https://doi.org/10.1103/PhysRevLett.86.3463} {\bibfield  {journal} {\bibinfo  {journal} {Phys. Rev. Lett.}\ }\textbf {\bibinfo {volume} {86}},\ \bibinfo {pages} {3463} (\bibinfo {year} {2001}{\natexlab{a}})}\BibitemShut {NoStop}%
\bibitem [{\citenamefont {Esposito}\ and\ \citenamefont {Van~den Broeck}(2010{\natexlab{a}})}]{Esposito2010dft}%
  \BibitemOpen
  \bibfield  {author} {\bibinfo {author} {\bibfnamefont {M.}~\bibnamefont {Esposito}}\ and\ \bibinfo {author} {\bibfnamefont {C.}~\bibnamefont {Van~den Broeck}},\ }\href {https://doi.org/10.1103/PhysRevLett.104.090601} {\bibfield  {journal} {\bibinfo  {journal} {Phys. Rev. Lett.}\ }\textbf {\bibinfo {volume} {104}},\ \bibinfo {pages} {090601} (\bibinfo {year} {2010}{\natexlab{a}})}\BibitemShut {NoStop}%
\bibitem [{\citenamefont {Horowitz}\ and\ \citenamefont {Sagawa}(2014)}]{horowitz_equivalent_2014}%
  \BibitemOpen
  \bibfield  {author} {\bibinfo {author} {\bibfnamefont {J.~M.}\ \bibnamefont {Horowitz}}\ and\ \bibinfo {author} {\bibfnamefont {T.}~\bibnamefont {Sagawa}},\ }\href {https://doi.org/10.1007/s10955-014-0991-1} {\bibfield  {journal} {\bibinfo  {journal} {Journal of Statistical Physics}\ }\textbf {\bibinfo {volume} {156}},\ \bibinfo {pages} {55} (\bibinfo {year} {2014})}\BibitemShut {NoStop}%
\bibitem [{\citenamefont {Manzano}\ \emph {et~al.}(2018)\citenamefont {Manzano}, \citenamefont {Horowitz},\ and\ \citenamefont {Parrondo}}]{manzano_quantum_2018}%
  \BibitemOpen
  \bibfield  {author} {\bibinfo {author} {\bibfnamefont {G.}~\bibnamefont {Manzano}}, \bibinfo {author} {\bibfnamefont {J.~M.}\ \bibnamefont {Horowitz}},\ and\ \bibinfo {author} {\bibfnamefont {J.~M.~R.}\ \bibnamefont {Parrondo}},\ }\href {https://doi.org/10.1103/PhysRevX.8.031037} {\bibfield  {journal} {\bibinfo  {journal} {Physical Review X}\ }\textbf {\bibinfo {volume} {8}},\ \bibinfo {pages} {031037} (\bibinfo {year} {2018})}\BibitemShut {NoStop}%
\bibitem [{\citenamefont {Mandal}\ and\ \citenamefont {Jarzynski}(2016)}]{mandal_analysis_2016}%
  \BibitemOpen
  \bibfield  {author} {\bibinfo {author} {\bibfnamefont {D.}~\bibnamefont {Mandal}}\ and\ \bibinfo {author} {\bibfnamefont {C.}~\bibnamefont {Jarzynski}},\ }\href {https://doi.org/10.1088/1742-5468/2016/06/063204} {\bibfield  {journal} {\bibinfo  {journal} {Journal of Statistical Mechanics: Theory and Experiment}\ }\textbf {\bibinfo {volume} {2016}},\ \bibinfo {pages} {063204} (\bibinfo {year} {2016})}\BibitemShut {NoStop}%
\bibitem [{\citenamefont {Acevedo}\ \emph {et~al.}(2014)\citenamefont {Acevedo}, \citenamefont {Quiroga}, \citenamefont {Rodr\'{\i}guez},\ and\ \citenamefont {Johnson}}]{Acevedo2014}%
  \BibitemOpen
  \bibfield  {author} {\bibinfo {author} {\bibfnamefont {O.~L.}\ \bibnamefont {Acevedo}}, \bibinfo {author} {\bibfnamefont {L.}~\bibnamefont {Quiroga}}, \bibinfo {author} {\bibfnamefont {F.~J.}\ \bibnamefont {Rodr\'{\i}guez}},\ and\ \bibinfo {author} {\bibfnamefont {N.~F.}\ \bibnamefont {Johnson}},\ }\href {https://doi.org/10.1103/PhysRevLett.112.030403} {\bibfield  {journal} {\bibinfo  {journal} {Phys. Rev. Lett.}\ }\textbf {\bibinfo {volume} {112}},\ \bibinfo {pages} {030403} (\bibinfo {year} {2014})}\BibitemShut {NoStop}%
\bibitem [{\citenamefont {Hwang}\ \emph {et~al.}(2015)\citenamefont {Hwang}, \citenamefont {Puebla},\ and\ \citenamefont {Plenio}}]{hwang_quantum_2015}%
  \BibitemOpen
  \bibfield  {author} {\bibinfo {author} {\bibfnamefont {M.-J.}\ \bibnamefont {Hwang}}, \bibinfo {author} {\bibfnamefont {R.}~\bibnamefont {Puebla}},\ and\ \bibinfo {author} {\bibfnamefont {M.~B.}\ \bibnamefont {Plenio}},\ }\href {https://doi.org/10.1103/PhysRevLett.115.180404} {\bibfield  {journal} {\bibinfo  {journal} {Physical Review Letters}\ }\textbf {\bibinfo {volume} {115}},\ \bibinfo {pages} {180404} (\bibinfo {year} {2015})}\BibitemShut {NoStop}%
\bibitem [{\citenamefont {De~Grandi}\ \emph {et~al.}(2010)\citenamefont {De~Grandi}, \citenamefont {Gritsev},\ and\ \citenamefont {Polkovnikov}}]{de_grandi_quench_2010}%
  \BibitemOpen
  \bibfield  {author} {\bibinfo {author} {\bibfnamefont {C.}~\bibnamefont {De~Grandi}}, \bibinfo {author} {\bibfnamefont {V.}~\bibnamefont {Gritsev}},\ and\ \bibinfo {author} {\bibfnamefont {A.}~\bibnamefont {Polkovnikov}},\ }\href {https://doi.org/10.1103/PhysRevB.81.012303} {\bibfield  {journal} {\bibinfo  {journal} {Physical Review B}\ }\textbf {\bibinfo {volume} {81}},\ \bibinfo {pages} {012303} (\bibinfo {year} {2010})}\BibitemShut {NoStop}%
\bibitem [{\citenamefont {Barankov}\ and\ \citenamefont {Polkovnikov}(2008)}]{Barankov_optimal_2008}%
  \BibitemOpen
  \bibfield  {author} {\bibinfo {author} {\bibfnamefont {R.}~\bibnamefont {Barankov}}\ and\ \bibinfo {author} {\bibfnamefont {A.}~\bibnamefont {Polkovnikov}},\ }\href {https://doi.org/10.1103/PhysRevLett.101.076801} {\bibfield  {journal} {\bibinfo  {journal} {Phys. Rev. Lett.}\ }\textbf {\bibinfo {volume} {101}},\ \bibinfo {pages} {076801} (\bibinfo {year} {2008})}\BibitemShut {NoStop}%
\bibitem [{\citenamefont {Cavina}\ \emph {et~al.}(2017)\citenamefont {Cavina}, \citenamefont {Mari},\ and\ \citenamefont {Giovannetti}}]{cavina_slow_2017}%
  \BibitemOpen
  \bibfield  {author} {\bibinfo {author} {\bibfnamefont {V.}~\bibnamefont {Cavina}}, \bibinfo {author} {\bibfnamefont {A.}~\bibnamefont {Mari}},\ and\ \bibinfo {author} {\bibfnamefont {V.}~\bibnamefont {Giovannetti}},\ }\href {https://doi.org/10.1103/PhysRevLett.119.050601} {\bibfield  {journal} {\bibinfo  {journal} {Physical Review Letters}\ }\textbf {\bibinfo {volume} {119}},\ \bibinfo {pages} {050601} (\bibinfo {year} {2017})}\BibitemShut {NoStop}%
\bibitem [{\citenamefont {Sivak}\ and\ \citenamefont {Crooks}(2012)}]{sivak_thermodynamic_2012}%
  \BibitemOpen
  \bibfield  {author} {\bibinfo {author} {\bibfnamefont {D.~A.}\ \bibnamefont {Sivak}}\ and\ \bibinfo {author} {\bibfnamefont {G.~E.}\ \bibnamefont {Crooks}},\ }\href {https://doi.org/10.1103/PhysRevLett.108.190602} {\bibfield  {journal} {\bibinfo  {journal} {Physical Review Letters}\ }\textbf {\bibinfo {volume} {108}},\ \bibinfo {pages} {190602} (\bibinfo {year} {2012})}\BibitemShut {NoStop}%
\bibitem [{\citenamefont {Crooks}(1999)}]{crooks_entropy_1999}%
  \BibitemOpen
  \bibfield  {author} {\bibinfo {author} {\bibfnamefont {G.~E.}\ \bibnamefont {Crooks}},\ }\href {https://doi.org/10.1103/PhysRevE.60.2721} {\bibfield  {journal} {\bibinfo  {journal} {Physical Review E}\ }\textbf {\bibinfo {volume} {60}},\ \bibinfo {pages} {2721} (\bibinfo {year} {1999})}\BibitemShut {NoStop}%
\bibitem [{\citenamefont {Scandi}\ and\ \citenamefont {Perarnau-Llobet}(2019)}]{scandi_thermodynamic_2019}%
  \BibitemOpen
  \bibfield  {author} {\bibinfo {author} {\bibfnamefont {M.}~\bibnamefont {Scandi}}\ and\ \bibinfo {author} {\bibfnamefont {M.}~\bibnamefont {Perarnau-Llobet}},\ }\href {https://doi.org/10.22331/q-2019-10-24-197} {\bibfield  {journal} {\bibinfo  {journal} {Quantum}\ }\textbf {\bibinfo {volume} {3}},\ \bibinfo {pages} {197} (\bibinfo {year} {2019})}\BibitemShut {NoStop}%
\bibitem [{\citenamefont {Kubo}\ \emph {et~al.}(2012)\citenamefont {Kubo}, \citenamefont {Toda},\ and\ \citenamefont {Hashitsume}}]{kubo_statistical_2012}%
  \BibitemOpen
  \bibfield  {author} {\bibinfo {author} {\bibfnamefont {R.}~\bibnamefont {Kubo}}, \bibinfo {author} {\bibfnamefont {M.}~\bibnamefont {Toda}},\ and\ \bibinfo {author} {\bibfnamefont {N.}~\bibnamefont {Hashitsume}},\ }\href@noop {} {\emph {\bibinfo {title} {Statistical Physics II: Nonequilibrium Statistical Mechanics}}}\ (\bibinfo  {publisher} {Springer Science \& Business Media},\ \bibinfo {year} {2012})\BibitemShut {NoStop}%
\bibitem [{Note1()}]{Note1}%
  \BibitemOpen
  \bibinfo {note} {To be explicit, we use Eq.~\protect \eqref {eq:sigma:na:dot} under the approximation that $\protect \hat {\rho }_t \approx e^{\protect \mathcal {L}_{g_c}(t-t^*)}\protect \hat {\rho }_{t^*}$ and $\protect \hat {\pi }_{g(t)} \approx \protect \hat {\pi }_{g_c}$. This is the standard formulation of entropy production for an evolution with fixed Lindblad generator $\protect \mathcal {L}_{g_c}$~\cite {spohnEntropyProductionQuantum1978}. Here $\protect \hat {\pi }_{g_c}$ should be interpreted in the limiting sense as $g_c$ is approached from below, since the steady state is, strictly speaking, ill-defined in the thermodynamic limit at $g = g_c$.}\BibitemShut {Stop}%
\bibitem [{\citenamefont {Stanley}(1999)}]{Stanley_scaling_1999}%
  \BibitemOpen
  \bibfield  {author} {\bibinfo {author} {\bibfnamefont {H.~E.}\ \bibnamefont {Stanley}},\ }\href {https://doi.org/10.1103/RevModPhys.71.S358} {\bibfield  {journal} {\bibinfo  {journal} {Rev. Mod. Phys.}\ }\textbf {\bibinfo {volume} {71}},\ \bibinfo {pages} {S358} (\bibinfo {year} {1999})}\BibitemShut {NoStop}%
\bibitem [{\citenamefont {Nagy}\ \emph {et~al.}(2011)\citenamefont {Nagy}, \citenamefont {Szirmai},\ and\ \citenamefont {Domokos}}]{nagy_critical_2011}%
  \BibitemOpen
  \bibfield  {author} {\bibinfo {author} {\bibfnamefont {D.}~\bibnamefont {Nagy}}, \bibinfo {author} {\bibfnamefont {G.}~\bibnamefont {Szirmai}},\ and\ \bibinfo {author} {\bibfnamefont {P.}~\bibnamefont {Domokos}},\ }\href {https://doi.org/10.1103/PhysRevA.84.043637} {\bibfield  {journal} {\bibinfo  {journal} {Physical Review A}\ }\textbf {\bibinfo {volume} {84}},\ \bibinfo {pages} {043637} (\bibinfo {year} {2011})}\BibitemShut {NoStop}%
\bibitem [{\citenamefont {Hwang}\ \emph {et~al.}(2018)\citenamefont {Hwang}, \citenamefont {Rabl},\ and\ \citenamefont {Plenio}}]{Hwang_dissipative_2018}%
  \BibitemOpen
  \bibfield  {author} {\bibinfo {author} {\bibfnamefont {M.-J.}\ \bibnamefont {Hwang}}, \bibinfo {author} {\bibfnamefont {P.}~\bibnamefont {Rabl}},\ and\ \bibinfo {author} {\bibfnamefont {M.~B.}\ \bibnamefont {Plenio}},\ }\href {https://doi.org/10.1103/PhysRevA.97.013825} {\bibfield  {journal} {\bibinfo  {journal} {Phys. Rev. A}\ }\textbf {\bibinfo {volume} {97}},\ \bibinfo {pages} {013825} (\bibinfo {year} {2018})}\BibitemShut {NoStop}%
\bibitem [{\citenamefont {Ferreira}\ and\ \citenamefont {Ribeiro}(2019)}]{Ferreira_LMG_2019}%
  \BibitemOpen
  \bibfield  {author} {\bibinfo {author} {\bibfnamefont {J.~a.~S.}\ \bibnamefont {Ferreira}}\ and\ \bibinfo {author} {\bibfnamefont {P.}~\bibnamefont {Ribeiro}},\ }\href {https://doi.org/10.1103/PhysRevB.100.184422} {\bibfield  {journal} {\bibinfo  {journal} {Phys. Rev. B}\ }\textbf {\bibinfo {volume} {100}},\ \bibinfo {pages} {184422} (\bibinfo {year} {2019})}\BibitemShut {NoStop}%
\bibitem [{\citenamefont {Dicke}(1954)}]{dicke_coherence_1954}%
  \BibitemOpen
  \bibfield  {author} {\bibinfo {author} {\bibfnamefont {R.~H.}\ \bibnamefont {Dicke}},\ }\href {https://doi.org/10.1103/PhysRev.93.99} {\bibfield  {journal} {\bibinfo  {journal} {Physical Review}\ }\textbf {\bibinfo {volume} {93}},\ \bibinfo {pages} {99} (\bibinfo {year} {1954})}\BibitemShut {NoStop}%
\bibitem [{\citenamefont {Hepp}\ and\ \citenamefont {Lieb}(1973)}]{hepp_equilibrium_1973}%
  \BibitemOpen
  \bibfield  {author} {\bibinfo {author} {\bibfnamefont {K.}~\bibnamefont {Hepp}}\ and\ \bibinfo {author} {\bibfnamefont {E.~H.}\ \bibnamefont {Lieb}},\ }\href {https://doi.org/10.1103/PhysRevA.8.2517} {\bibfield  {journal} {\bibinfo  {journal} {Physical Review A}\ }\textbf {\bibinfo {volume} {8}},\ \bibinfo {pages} {2517} (\bibinfo {year} {1973})}\BibitemShut {NoStop}%
\bibitem [{\citenamefont {Wang}\ and\ \citenamefont {Hioe}(1973)}]{wang_phase_1973}%
  \BibitemOpen
  \bibfield  {author} {\bibinfo {author} {\bibfnamefont {Y.~K.}\ \bibnamefont {Wang}}\ and\ \bibinfo {author} {\bibfnamefont {F.~T.}\ \bibnamefont {Hioe}},\ }\href {https://doi.org/10.1103/PhysRevA.7.831} {\bibfield  {journal} {\bibinfo  {journal} {Physical Review A}\ }\textbf {\bibinfo {volume} {7}},\ \bibinfo {pages} {831} (\bibinfo {year} {1973})}\BibitemShut {NoStop}%
\bibitem [{\citenamefont {Hioe}(1973)}]{hioe_phase_1973}%
  \BibitemOpen
  \bibfield  {author} {\bibinfo {author} {\bibfnamefont {F.~T.}\ \bibnamefont {Hioe}},\ }\href {https://doi.org/10.1103/PhysRevA.8.1440} {\bibfield  {journal} {\bibinfo  {journal} {Physical Review A}\ }\textbf {\bibinfo {volume} {8}},\ \bibinfo {pages} {1440} (\bibinfo {year} {1973})}\BibitemShut {NoStop}%
\bibitem [{\citenamefont {Lang}\ and\ \citenamefont {Piazza}(2016)}]{lang_critical_2016}%
  \BibitemOpen
  \bibfield  {author} {\bibinfo {author} {\bibfnamefont {J.}~\bibnamefont {Lang}}\ and\ \bibinfo {author} {\bibfnamefont {F.}~\bibnamefont {Piazza}},\ }\href {https://doi.org/10.1103/PhysRevA.94.033628} {\bibfield  {journal} {\bibinfo  {journal} {Physical Review A}\ }\textbf {\bibinfo {volume} {94}},\ \bibinfo {pages} {033628} (\bibinfo {year} {2016})}\BibitemShut {NoStop}%
\bibitem [{\citenamefont {Paz}\ and\ \citenamefont {Maghrebi}(2022)}]{paz_driven-dissipative_2022}%
  \BibitemOpen
  \bibfield  {author} {\bibinfo {author} {\bibfnamefont {D.~A.}\ \bibnamefont {Paz}}\ and\ \bibinfo {author} {\bibfnamefont {M.~F.}\ \bibnamefont {Maghrebi}},\ }\href {https://doi.org/10.1209/0295-5075/ac33cb} {\bibfield  {journal} {\bibinfo  {journal} {Europhysics Letters}\ }\textbf {\bibinfo {volume} {136}},\ \bibinfo {pages} {10002} (\bibinfo {year} {2022})}\BibitemShut {NoStop}%
\bibitem [{\citenamefont {Dimer}\ \emph {et~al.}(2007)\citenamefont {Dimer}, \citenamefont {Estienne}, \citenamefont {Parkins},\ and\ \citenamefont {Carmichael}}]{dimer_proposed_2007}%
  \BibitemOpen
  \bibfield  {author} {\bibinfo {author} {\bibfnamefont {F.}~\bibnamefont {Dimer}}, \bibinfo {author} {\bibfnamefont {B.}~\bibnamefont {Estienne}}, \bibinfo {author} {\bibfnamefont {A.~S.}\ \bibnamefont {Parkins}},\ and\ \bibinfo {author} {\bibfnamefont {H.~J.}\ \bibnamefont {Carmichael}},\ }\href {https://doi.org/10.1103/PhysRevA.75.013804} {\bibfield  {journal} {\bibinfo  {journal} {Physical Review A}\ }\textbf {\bibinfo {volume} {75}},\ \bibinfo {pages} {013804} (\bibinfo {year} {2007})}\BibitemShut {NoStop}%
\bibitem [{\citenamefont {\"Oztop}\ \emph {et~al.}(2012)\citenamefont {\"Oztop}, \citenamefont {Bordyuh}, \citenamefont {M\"ustecaplioglu},\ and\ \citenamefont {T\"ureci}}]{oztop_excitations_2012}%
  \BibitemOpen
  \bibfield  {author} {\bibinfo {author} {\bibfnamefont {B.}~\bibnamefont {\"Oztop}}, \bibinfo {author} {\bibfnamefont {M.}~\bibnamefont {Bordyuh}}, \bibinfo {author} {\bibfnamefont {O.~E.}\ \bibnamefont {M\"ustecaplioglu}},\ and\ \bibinfo {author} {\bibfnamefont {H.~E.}\ \bibnamefont {T\"ureci}},\ }\href {https://doi.org/10.1088/1367-2630/14/8/085011} {\bibfield  {journal} {\bibinfo  {journal} {New Journal of Physics}\ }\textbf {\bibinfo {volume} {14}},\ \bibinfo {pages} {085011} (\bibinfo {year} {2012})}\BibitemShut {NoStop}%
\bibitem [{\citenamefont {Kirton}\ \emph {et~al.}(2019)\citenamefont {Kirton}, \citenamefont {Roses}, \citenamefont {Keeling},\ and\ \citenamefont {Torre}}]{kirton_introduction_2019}%
  \BibitemOpen
  \bibfield  {author} {\bibinfo {author} {\bibfnamefont {P.}~\bibnamefont {Kirton}}, \bibinfo {author} {\bibfnamefont {M.~M.}\ \bibnamefont {Roses}}, \bibinfo {author} {\bibfnamefont {J.}~\bibnamefont {Keeling}},\ and\ \bibinfo {author} {\bibfnamefont {E.~G.~D.}\ \bibnamefont {Torre}},\ }\href {https://doi.org/10.1002/qute.201800043} {\bibfield  {journal} {\bibinfo  {journal} {Advanced Quantum Technologies}\ }\textbf {\bibinfo {volume} {2}},\ \bibinfo {pages} {1800043} (\bibinfo {year} {2019})}\BibitemShut {NoStop}%
\bibitem [{\citenamefont {Ruppeiner}(1995)}]{Ruppeiner_Riemannian_1995}%
  \BibitemOpen
  \bibfield  {author} {\bibinfo {author} {\bibfnamefont {G.}~\bibnamefont {Ruppeiner}},\ }\href {https://doi.org/10.1103/RevModPhys.67.605} {\bibfield  {journal} {\bibinfo  {journal} {Rev. Mod. Phys.}\ }\textbf {\bibinfo {volume} {67}},\ \bibinfo {pages} {605} (\bibinfo {year} {1995})}\BibitemShut {NoStop}%
\bibitem [{\citenamefont {Brody}\ and\ \citenamefont {Rivier}(1995)}]{Brody_Geometrical_1995}%
  \BibitemOpen
  \bibfield  {author} {\bibinfo {author} {\bibfnamefont {D.}~\bibnamefont {Brody}}\ and\ \bibinfo {author} {\bibfnamefont {N.}~\bibnamefont {Rivier}},\ }\href {https://doi.org/10.1103/PhysRevE.51.1006} {\bibfield  {journal} {\bibinfo  {journal} {Phys. Rev. E}\ }\textbf {\bibinfo {volume} {51}},\ \bibinfo {pages} {1006} (\bibinfo {year} {1995})}\BibitemShut {NoStop}%
\bibitem [{\citenamefont {Zanardi}\ and\ \citenamefont {Paunković}(2006)}]{zanardi_ground_2006}%
  \BibitemOpen
  \bibfield  {author} {\bibinfo {author} {\bibfnamefont {P.}~\bibnamefont {Zanardi}}\ and\ \bibinfo {author} {\bibfnamefont {N.}~\bibnamefont {Paunković}},\ }\href {https://doi.org/10.1103/PhysRevE.74.031123} {\bibfield  {journal} {\bibinfo  {journal} {Physical Review E}\ }\textbf {\bibinfo {volume} {74}},\ \bibinfo {pages} {031123} (\bibinfo {year} {2006})}\BibitemShut {NoStop}%
\bibitem [{\citenamefont {Venuti}\ and\ \citenamefont {Zanardi}(2007)}]{venuti_quantum_2007}%
  \BibitemOpen
  \bibfield  {author} {\bibinfo {author} {\bibfnamefont {L.~C.}\ \bibnamefont {Venuti}}\ and\ \bibinfo {author} {\bibfnamefont {P.}~\bibnamefont {Zanardi}},\ }\href {https://doi.org/10.1103/PhysRevLett.99.095701} {\bibfield  {journal} {\bibinfo  {journal} {Physical Review Letters}\ }\textbf {\bibinfo {volume} {99}},\ \bibinfo {pages} {095701} (\bibinfo {year} {2007})}\BibitemShut {NoStop}%
\bibitem [{\citenamefont {You}\ \emph {et~al.}(2007)\citenamefont {You}, \citenamefont {Li},\ and\ \citenamefont {Gu}}]{you_fidelity_2007}%
  \BibitemOpen
  \bibfield  {author} {\bibinfo {author} {\bibfnamefont {W.-L.}\ \bibnamefont {You}}, \bibinfo {author} {\bibfnamefont {Y.-W.}\ \bibnamefont {Li}},\ and\ \bibinfo {author} {\bibfnamefont {S.-J.}\ \bibnamefont {Gu}},\ }\href {https://doi.org/10.1103/PhysRevE.76.022101} {\bibfield  {journal} {\bibinfo  {journal} {Physical Review E}\ }\textbf {\bibinfo {volume} {76}},\ \bibinfo {pages} {022101} (\bibinfo {year} {2007})}\BibitemShut {NoStop}%
\bibitem [{\citenamefont {Zanardi}\ \emph {et~al.}(2007)\citenamefont {Zanardi}, \citenamefont {Giorda},\ and\ \citenamefont {Cozzini}}]{zanardi_information-theoretic_2007}%
  \BibitemOpen
  \bibfield  {author} {\bibinfo {author} {\bibfnamefont {P.}~\bibnamefont {Zanardi}}, \bibinfo {author} {\bibfnamefont {P.}~\bibnamefont {Giorda}},\ and\ \bibinfo {author} {\bibfnamefont {M.}~\bibnamefont {Cozzini}},\ }\href {https://doi.org/10.1103/PhysRevLett.99.100603} {\bibfield  {journal} {\bibinfo  {journal} {Physical Review Letters}\ }\textbf {\bibinfo {volume} {99}},\ \bibinfo {pages} {100603} (\bibinfo {year} {2007})}\BibitemShut {NoStop}%
\bibitem [{\citenamefont {Banchi}\ \emph {et~al.}(2014)\citenamefont {Banchi}, \citenamefont {Giorda},\ and\ \citenamefont {Zanardi}}]{banchi_quantum_2014}%
  \BibitemOpen
  \bibfield  {author} {\bibinfo {author} {\bibfnamefont {L.}~\bibnamefont {Banchi}}, \bibinfo {author} {\bibfnamefont {P.}~\bibnamefont {Giorda}},\ and\ \bibinfo {author} {\bibfnamefont {P.}~\bibnamefont {Zanardi}},\ }\href {https://doi.org/10.1103/PhysRevE.89.022102} {\bibfield  {journal} {\bibinfo  {journal} {Physical Review E}\ }\textbf {\bibinfo {volume} {89}},\ \bibinfo {pages} {022102} (\bibinfo {year} {2014})}\BibitemShut {NoStop}%
\bibitem [{\citenamefont {Wang}\ \emph {et~al.}(2014)\citenamefont {Wang}, \citenamefont {Wu}, \citenamefont {Yang}, \citenamefont {Jin}, \citenamefont {Lambert},\ and\ \citenamefont {Nori}}]{wang_quantum_2014}%
  \BibitemOpen
  \bibfield  {author} {\bibinfo {author} {\bibfnamefont {T.~L.}\ \bibnamefont {Wang}}, \bibinfo {author} {\bibfnamefont {L.~N.}\ \bibnamefont {Wu}}, \bibinfo {author} {\bibfnamefont {W.}~\bibnamefont {Yang}}, \bibinfo {author} {\bibfnamefont {G.~R.}\ \bibnamefont {Jin}}, \bibinfo {author} {\bibfnamefont {N.}~\bibnamefont {Lambert}},\ and\ \bibinfo {author} {\bibfnamefont {F.}~\bibnamefont {Nori}},\ }\href {https://doi.org/10.1088/1367-2630/16/6/063039} {\bibfield  {journal} {\bibinfo  {journal} {New Journal of Physics}\ }\textbf {\bibinfo {volume} {16}},\ \bibinfo {pages} {063039} (\bibinfo {year} {2014})}\BibitemShut {NoStop}%
\bibitem [{\citenamefont {Marzolino}\ and\ \citenamefont {Prosen}(2017)}]{Marzolino_Fisher_2017}%
  \BibitemOpen
  \bibfield  {author} {\bibinfo {author} {\bibfnamefont {U.}~\bibnamefont {Marzolino}}\ and\ \bibinfo {author} {\bibfnamefont {T.~c.~v.}\ \bibnamefont {Prosen}},\ }\href {https://doi.org/10.1103/PhysRevB.96.104402} {\bibfield  {journal} {\bibinfo  {journal} {Phys. Rev. B}\ }\textbf {\bibinfo {volume} {96}},\ \bibinfo {pages} {104402} (\bibinfo {year} {2017})}\BibitemShut {NoStop}%
\bibitem [{\citenamefont {Liu}\ \emph {et~al.}(2025)\citenamefont {Liu}, \citenamefont {Nian}, \citenamefont {Zhou}, \citenamefont {Xiong}, \citenamefont {L\"u},\ and\ \citenamefont {Zheng}}]{Liu_universal_2025}%
  \BibitemOpen
  \bibfield  {author} {\bibinfo {author} {\bibfnamefont {J.-Y.}\ \bibnamefont {Liu}}, \bibinfo {author} {\bibfnamefont {L.-L.}\ \bibnamefont {Nian}}, \bibinfo {author} {\bibfnamefont {N.}~\bibnamefont {Zhou}}, \bibinfo {author} {\bibfnamefont {L.}~\bibnamefont {Xiong}}, \bibinfo {author} {\bibfnamefont {J.-T.}\ \bibnamefont {L\"u}},\ and\ \bibinfo {author} {\bibfnamefont {B.}~\bibnamefont {Zheng}},\ }\href {https://doi.org/10.1103/PhysRevA.111.L040201} {\bibfield  {journal} {\bibinfo  {journal} {Phys. Rev. A}\ }\textbf {\bibinfo {volume} {111}},\ \bibinfo {pages} {L040201} (\bibinfo {year} {2025})}\BibitemShut {NoStop}%
\bibitem [{\citenamefont {Trepagnier}\ \emph {et~al.}(2004)\citenamefont {Trepagnier}, \citenamefont {Jarzynski}, \citenamefont {Ritort}, \citenamefont {Crooks}, \citenamefont {Bustamante},\ and\ \citenamefont {Liphardt}}]{Trepagnier2004}%
  \BibitemOpen
  \bibfield  {author} {\bibinfo {author} {\bibfnamefont {E.~H.}\ \bibnamefont {Trepagnier}}, \bibinfo {author} {\bibfnamefont {C.}~\bibnamefont {Jarzynski}}, \bibinfo {author} {\bibfnamefont {F.}~\bibnamefont {Ritort}}, \bibinfo {author} {\bibfnamefont {G.~E.}\ \bibnamefont {Crooks}}, \bibinfo {author} {\bibfnamefont {C.~J.}\ \bibnamefont {Bustamante}},\ and\ \bibinfo {author} {\bibfnamefont {J.}~\bibnamefont {Liphardt}},\ }\href {https://doi.org/10.1073/pnas.0406405101} {\bibfield  {journal} {\bibinfo  {journal} {Proceedings of the National Academy of Sciences}\ }\textbf {\bibinfo {volume} {101}},\ \bibinfo {pages} {15038–15041} (\bibinfo {year} {2004})}\BibitemShut {NoStop}%
\bibitem [{\citenamefont {Mounier}\ and\ \citenamefont {Naert}(2012)}]{Mounier2012}%
  \BibitemOpen
  \bibfield  {author} {\bibinfo {author} {\bibfnamefont {A.}~\bibnamefont {Mounier}}\ and\ \bibinfo {author} {\bibfnamefont {A.}~\bibnamefont {Naert}},\ }\href {https://doi.org/10.1209/0295-5075/100/30002} {\bibfield  {journal} {\bibinfo  {journal} {EPL (Europhysics Letters)}\ }\textbf {\bibinfo {volume} {100}},\ \bibinfo {pages} {30002} (\bibinfo {year} {2012})}\BibitemShut {NoStop}%
\bibitem [{\citenamefont {Granger}\ \emph {et~al.}(2015)\citenamefont {Granger}, \citenamefont {Mehlis}, \citenamefont {Rold\'{a}n}, \citenamefont {Ciliberto},\ and\ \citenamefont {Kantz}}]{Granger2015}%
  \BibitemOpen
  \bibfield  {author} {\bibinfo {author} {\bibfnamefont {L.}~\bibnamefont {Granger}}, \bibinfo {author} {\bibfnamefont {J.}~\bibnamefont {Mehlis}}, \bibinfo {author} {\bibfnamefont {E.}~\bibnamefont {Rold\'{a}n}}, \bibinfo {author} {\bibfnamefont {S.}~\bibnamefont {Ciliberto}},\ and\ \bibinfo {author} {\bibfnamefont {H.}~\bibnamefont {Kantz}},\ }\href {https://doi.org/10.1088/1367-2630/17/6/065005} {\bibfield  {journal} {\bibinfo  {journal} {New Journal of Physics}\ }\textbf {\bibinfo {volume} {17}},\ \bibinfo {pages} {065005} (\bibinfo {year} {2015})}\BibitemShut {NoStop}%
\bibitem [{\citenamefont {Joubaud}\ \emph {et~al.}(2008)\citenamefont {Joubaud}, \citenamefont {Garnier},\ and\ \citenamefont {Ciliberto}}]{Joubaud2008}%
  \BibitemOpen
  \bibfield  {author} {\bibinfo {author} {\bibfnamefont {S.}~\bibnamefont {Joubaud}}, \bibinfo {author} {\bibfnamefont {N.~B.}\ \bibnamefont {Garnier}},\ and\ \bibinfo {author} {\bibfnamefont {S.}~\bibnamefont {Ciliberto}},\ }\href {https://doi.org/10.1209/0295-5075/82/30007} {\bibfield  {journal} {\bibinfo  {journal} {EPL (Europhysics Letters)}\ }\textbf {\bibinfo {volume} {82}},\ \bibinfo {pages} {30007} (\bibinfo {year} {2008})}\BibitemShut {NoStop}%
\bibitem [{\citenamefont {Wang}\ \emph {et~al.}(2002)\citenamefont {Wang}, \citenamefont {Sevick}, \citenamefont {Mittag}, \citenamefont {Searles},\ and\ \citenamefont {Evans}}]{Wang2002}%
  \BibitemOpen
  \bibfield  {author} {\bibinfo {author} {\bibfnamefont {G.~M.}\ \bibnamefont {Wang}}, \bibinfo {author} {\bibfnamefont {E.~M.}\ \bibnamefont {Sevick}}, \bibinfo {author} {\bibfnamefont {E.}~\bibnamefont {Mittag}}, \bibinfo {author} {\bibfnamefont {D.~J.}\ \bibnamefont {Searles}},\ and\ \bibinfo {author} {\bibfnamefont {D.~J.}\ \bibnamefont {Evans}},\ }\bibfield  {journal} {\bibinfo  {journal} {Physical Review Letters}\ }\textbf {\bibinfo {volume} {89}},\ \href {https://doi.org/10.1103/physrevlett.89.050601} {10.1103/physrevlett.89.050601} (\bibinfo {year} {2002})\BibitemShut {NoStop}%
\bibitem [{\citenamefont {Hatano}\ and\ \citenamefont {Sasa}(2001{\natexlab{b}})}]{hatano_steady-state_2001}%
  \BibitemOpen
  \bibfield  {author} {\bibinfo {author} {\bibfnamefont {T.}~\bibnamefont {Hatano}}\ and\ \bibinfo {author} {\bibfnamefont {S.-i.}\ \bibnamefont {Sasa}},\ }\href {https://doi.org/10.1103/PhysRevLett.86.3463} {\bibfield  {journal} {\bibinfo  {journal} {Physical Review Letters}\ }\textbf {\bibinfo {volume} {86}},\ \bibinfo {pages} {3463} (\bibinfo {year} {2001}{\natexlab{b}})}\BibitemShut {NoStop}%
\bibitem [{\citenamefont {Esposito}\ and\ \citenamefont {Van~den Broeck}(2010{\natexlab{b}})}]{esposito_three_2010}%
  \BibitemOpen
  \bibfield  {author} {\bibinfo {author} {\bibfnamefont {M.}~\bibnamefont {Esposito}}\ and\ \bibinfo {author} {\bibfnamefont {C.}~\bibnamefont {Van~den Broeck}},\ }\href {https://doi.org/10.1103/PhysRevE.82.011143} {\bibfield  {journal} {\bibinfo  {journal} {Physical Review E}\ }\textbf {\bibinfo {volume} {82}},\ \bibinfo {pages} {011143} (\bibinfo {year} {2010}{\natexlab{b}})}\BibitemShut {NoStop}%
\bibitem [{\citenamefont {Koski}\ \emph {et~al.}(2013)\citenamefont {Koski}, \citenamefont {Sagawa}, \citenamefont {Saira}, \citenamefont {Yoon}, \citenamefont {Kutvonen}, \citenamefont {Solinas}, \citenamefont {M\"{o}tt\"{o}nen}, \citenamefont {Ala-Nissila},\ and\ \citenamefont {Pekola}}]{Koski2013}%
  \BibitemOpen
  \bibfield  {author} {\bibinfo {author} {\bibfnamefont {J.~V.}\ \bibnamefont {Koski}}, \bibinfo {author} {\bibfnamefont {T.}~\bibnamefont {Sagawa}}, \bibinfo {author} {\bibfnamefont {O.-P.}\ \bibnamefont {Saira}}, \bibinfo {author} {\bibfnamefont {Y.}~\bibnamefont {Yoon}}, \bibinfo {author} {\bibfnamefont {A.}~\bibnamefont {Kutvonen}}, \bibinfo {author} {\bibfnamefont {P.}~\bibnamefont {Solinas}}, \bibinfo {author} {\bibfnamefont {M.}~\bibnamefont {M\"{o}tt\"{o}nen}}, \bibinfo {author} {\bibfnamefont {T.}~\bibnamefont {Ala-Nissila}},\ and\ \bibinfo {author} {\bibfnamefont {J.~P.}\ \bibnamefont {Pekola}},\ }\href {https://doi.org/10.1038/nphys2711} {\bibfield  {journal} {\bibinfo  {journal} {Nature Physics}\ }\textbf {\bibinfo {volume} {9}},\ \bibinfo {pages} {644–648} (\bibinfo {year} {2013})}\BibitemShut {NoStop}%
\bibitem [{\citenamefont {Hekking}\ and\ \citenamefont {Pekola}(2013)}]{Hekking2013}%
  \BibitemOpen
  \bibfield  {author} {\bibinfo {author} {\bibfnamefont {F.~W.~J.}\ \bibnamefont {Hekking}}\ and\ \bibinfo {author} {\bibfnamefont {J.~P.}\ \bibnamefont {Pekola}},\ }\href@noop {} {\bibfield  {journal} {\bibinfo  {journal} {Phys. Rev. Lett.}\ }\textbf {\bibinfo {volume} {111}},\ \bibinfo {pages} {093602} (\bibinfo {year} {2013})}\BibitemShut {NoStop}%
\bibitem [{\citenamefont {Pekola}(2015)}]{pekola_towards_2015}%
  \BibitemOpen
  \bibfield  {author} {\bibinfo {author} {\bibfnamefont {J.~P.}\ \bibnamefont {Pekola}},\ }\href {https://doi.org/10.1038/nphys3169} {\bibfield  {journal} {\bibinfo  {journal} {Nature Physics}\ }\textbf {\bibinfo {volume} {11}},\ \bibinfo {pages} {118} (\bibinfo {year} {2015})}\BibitemShut {NoStop}%
\bibitem [{\citenamefont {Rossi}\ \emph {et~al.}(2020)\citenamefont {Rossi}, \citenamefont {Mancino}, \citenamefont {Landi}, \citenamefont {Paternostro}, \citenamefont {Schliesser},\ and\ \citenamefont {Belenchia}}]{Rossi2020}%
  \BibitemOpen
  \bibfield  {author} {\bibinfo {author} {\bibfnamefont {M.}~\bibnamefont {Rossi}}, \bibinfo {author} {\bibfnamefont {L.}~\bibnamefont {Mancino}}, \bibinfo {author} {\bibfnamefont {G.~T.}\ \bibnamefont {Landi}}, \bibinfo {author} {\bibfnamefont {M.}~\bibnamefont {Paternostro}}, \bibinfo {author} {\bibfnamefont {A.}~\bibnamefont {Schliesser}},\ and\ \bibinfo {author} {\bibfnamefont {A.}~\bibnamefont {Belenchia}},\ }\href {http://dx.doi.org/10.1103/PHYSREVLETT.125.080601} {\bibfield  {journal} {\bibinfo  {journal} {Physical Review Letters}\ }\textbf {\bibinfo {volume} {125}} (\bibinfo {year} {2020})}\BibitemShut {NoStop}%
\bibitem [{\citenamefont {Aguilar}\ \emph {et~al.}(2022)\citenamefont {Aguilar}, \citenamefont {Silva}, \citenamefont {Guimarães}, \citenamefont {Piera}, \citenamefont {Céleri},\ and\ \citenamefont {Landi}}]{Aguilar2022}%
  \BibitemOpen
  \bibfield  {author} {\bibinfo {author} {\bibfnamefont {G.~H.}\ \bibnamefont {Aguilar}}, \bibinfo {author} {\bibfnamefont {T.~L.}\ \bibnamefont {Silva}}, \bibinfo {author} {\bibfnamefont {T.~E.}\ \bibnamefont {Guimarães}}, \bibinfo {author} {\bibfnamefont {R.~S.}\ \bibnamefont {Piera}}, \bibinfo {author} {\bibfnamefont {L.~C.}\ \bibnamefont {Céleri}},\ and\ \bibinfo {author} {\bibfnamefont {G.~T.}\ \bibnamefont {Landi}},\ }\href {http://dx.doi.org/10.1103/PhysRevA.106.L020201} {\bibfield  {journal} {\bibinfo  {journal} {Physical Review A}\ }\textbf {\bibinfo {volume} {106}} (\bibinfo {year} {2022})}\BibitemShut {NoStop}%
\bibitem [{\citenamefont {Aamir}\ \emph {et~al.}(2025)\citenamefont {Aamir}, \citenamefont {Jamet~Suria}, \citenamefont {Mar{\'i}n~Guzm{\'a}n}, \citenamefont {{Castillo-Moreno}}, \citenamefont {Epstein}, \citenamefont {Yunger~Halpern},\ and\ \citenamefont {Gasparinetti}}]{aamirThermallyDrivenQuantum2025}%
  \BibitemOpen
  \bibfield  {author} {\bibinfo {author} {\bibfnamefont {M.~A.}\ \bibnamefont {Aamir}}, \bibinfo {author} {\bibfnamefont {P.}~\bibnamefont {Jamet~Suria}}, \bibinfo {author} {\bibfnamefont {J.~A.}\ \bibnamefont {Mar{\'i}n~Guzm{\'a}n}}, \bibinfo {author} {\bibfnamefont {C.}~\bibnamefont {{Castillo-Moreno}}}, \bibinfo {author} {\bibfnamefont {J.~M.}\ \bibnamefont {Epstein}}, \bibinfo {author} {\bibfnamefont {N.}~\bibnamefont {Yunger~Halpern}},\ and\ \bibinfo {author} {\bibfnamefont {S.}~\bibnamefont {Gasparinetti}},\ }\href {https://doi.org/10.1038/s41567-024-02708-5} {\bibfield  {journal} {\bibinfo  {journal} {Nature Physics}\ }\textbf {\bibinfo {volume} {21}},\ \bibinfo {pages} {1} (\bibinfo {year} {2025})}\BibitemShut {NoStop}%
\bibitem [{\citenamefont {Wadhia}\ \emph {et~al.}(2025)\citenamefont {Wadhia}, \citenamefont {Meier}, \citenamefont {Fedele}, \citenamefont {Silva}, \citenamefont {Nurgalieva}, \citenamefont {Craig}, \citenamefont {Jirovec}, \citenamefont {Saez-Mollejo}, \citenamefont {Ballabio}, \citenamefont {Chrastina}, \citenamefont {Isella}, \citenamefont {Huber}, \citenamefont {Mitchison}, \citenamefont {Erker},\ and\ \citenamefont {Ares}}]{Wadhia2025}%
  \BibitemOpen
  \bibfield  {author} {\bibinfo {author} {\bibfnamefont {V.}~\bibnamefont {Wadhia}}, \bibinfo {author} {\bibfnamefont {F.}~\bibnamefont {Meier}}, \bibinfo {author} {\bibfnamefont {F.}~\bibnamefont {Fedele}}, \bibinfo {author} {\bibfnamefont {R.}~\bibnamefont {Silva}}, \bibinfo {author} {\bibfnamefont {N.}~\bibnamefont {Nurgalieva}}, \bibinfo {author} {\bibfnamefont {D.~L.}\ \bibnamefont {Craig}}, \bibinfo {author} {\bibfnamefont {D.}~\bibnamefont {Jirovec}}, \bibinfo {author} {\bibfnamefont {J.}~\bibnamefont {Saez-Mollejo}}, \bibinfo {author} {\bibfnamefont {A.}~\bibnamefont {Ballabio}}, \bibinfo {author} {\bibfnamefont {D.}~\bibnamefont {Chrastina}}, \bibinfo {author} {\bibfnamefont {G.}~\bibnamefont {Isella}}, \bibinfo {author} {\bibfnamefont {M.}~\bibnamefont {Huber}}, \bibinfo {author} {\bibfnamefont {M.~T.}\ \bibnamefont {Mitchison}}, \bibinfo {author} {\bibfnamefont {P.}~\bibnamefont {Erker}},\ and\ \bibinfo {author} {\bibfnamefont {N.}~\bibnamefont {Ares}},\ }\href {https://doi.org/10.1103/5rtj-djfk}
  {\bibfield  {journal} {\bibinfo  {journal} {Phys. Rev. Lett.}\ }\textbf {\bibinfo {volume} {135}},\ \bibinfo {pages} {200407} (\bibinfo {year} {2025})}\BibitemShut {NoStop}%
\bibitem [{\citenamefont {{Bettmann, L. P. \emph{et al.}}}()}]{bettmann_inprep}%
  \BibitemOpen
  \bibfield  {author} {\bibinfo {author} {\bibnamefont {{Bettmann, L. P. \emph{et al.}}}},\ }\bibinfo {note} {in preparation}\BibitemShut {NoStop}%
\bibitem [{\citenamefont {Goes}\ \emph {et~al.}(2020)\citenamefont {Goes}, \citenamefont {Fiore},\ and\ \citenamefont {Landi}}]{goes_quantum_2020}%
  \BibitemOpen
  \bibfield  {author} {\bibinfo {author} {\bibfnamefont {B.~O.}\ \bibnamefont {Goes}}, \bibinfo {author} {\bibfnamefont {C.~E.}\ \bibnamefont {Fiore}},\ and\ \bibinfo {author} {\bibfnamefont {G.~T.}\ \bibnamefont {Landi}},\ }\href {https://doi.org/10.1103/PhysRevResearch.2.013136} {\bibfield  {journal} {\bibinfo  {journal} {Physical Review Research}\ }\textbf {\bibinfo {volume} {2}},\ \bibinfo {pages} {013136} (\bibinfo {year} {2020})}\BibitemShut {NoStop}%
\bibitem [{\citenamefont {Spohn}(1978)}]{spohnEntropyProductionQuantum1978}%
  \BibitemOpen
  \bibfield  {author} {\bibinfo {author} {\bibfnamefont {H.}~\bibnamefont {Spohn}},\ }\href {https://doi.org/10.1063/1.523789} {\bibfield  {journal} {\bibinfo  {journal} {Journal of Mathematical Physics}\ }\textbf {\bibinfo {volume} {19}},\ \bibinfo {pages} {1227} (\bibinfo {year} {1978})}\BibitemShut {NoStop}%
\bibitem [{\citenamefont {Gower}\ \emph {et~al.}(1972)\citenamefont {Gower}, \citenamefont {Boullion},\ and\ \citenamefont {Odell}}]{Gower1972}%
  \BibitemOpen
  \bibfield  {author} {\bibinfo {author} {\bibfnamefont {J.~C.}\ \bibnamefont {Gower}}, \bibinfo {author} {\bibfnamefont {T.~L.}\ \bibnamefont {Boullion}},\ and\ \bibinfo {author} {\bibfnamefont {P.~L.}\ \bibnamefont {Odell}},\ }\href@noop {} {\bibfield  {journal} {\bibinfo  {journal} {Technometrics}\ }\textbf {\bibinfo {volume} {14}},\ \bibinfo {pages} {806} (\bibinfo {year} {1972})}\BibitemShut {NoStop}%
\bibitem [{\citenamefont {Landi}\ \emph {et~al.}(2024)\citenamefont {Landi}, \citenamefont {Kewming}, \citenamefont {Mitchison},\ and\ \citenamefont {Potts}}]{landi_current_2024}%
  \BibitemOpen
  \bibfield  {author} {\bibinfo {author} {\bibfnamefont {G.~T.}\ \bibnamefont {Landi}}, \bibinfo {author} {\bibfnamefont {M.~J.}\ \bibnamefont {Kewming}}, \bibinfo {author} {\bibfnamefont {M.~T.}\ \bibnamefont {Mitchison}},\ and\ \bibinfo {author} {\bibfnamefont {P.~P.}\ \bibnamefont {Potts}},\ }\href {https://doi.org/10.1103/PRXQuantum.5.020201} {\bibfield  {journal} {\bibinfo  {journal} {PRX Quantum}\ }\textbf {\bibinfo {volume} {5}},\ \bibinfo {pages} {020201} (\bibinfo {year} {2024})}\BibitemShut {NoStop}%
\bibitem [{\citenamefont {Serafini}(2023)}]{Serafini2023-er}%
  \BibitemOpen
  \bibfield  {author} {\bibinfo {author} {\bibfnamefont {A.}~\bibnamefont {Serafini}},\ }\href@noop {} {\emph {\bibinfo {title} {Quantum continuous variables}}},\ \bibinfo {edition} {2nd}\ ed.\ (\bibinfo  {publisher} {CRC Press},\ \bibinfo {address} {London, England},\ \bibinfo {year} {2023})\BibitemShut {NoStop}%
\bibitem [{\citenamefont {Prosen}\ and\ \citenamefont {Seligman}(2010)}]{prosen_quantization_2010}%
  \BibitemOpen
  \bibfield  {author} {\bibinfo {author} {\bibfnamefont {T.}~\bibnamefont {Prosen}}\ and\ \bibinfo {author} {\bibfnamefont {T.~H.}\ \bibnamefont {Seligman}},\ }\href {https://doi.org/10.1088/1751-8113/43/39/392004} {\bibfield  {journal} {\bibinfo  {journal} {Journal of Physics A: Mathematical and Theoretical}\ }\textbf {\bibinfo {volume} {43}},\ \bibinfo {pages} {392004} (\bibinfo {year} {2010})}\BibitemShut {NoStop}%
\bibitem [{\citenamefont {Barthel}\ and\ \citenamefont {Zhang}(2022)}]{barthel_solving_2022}%
  \BibitemOpen
  \bibfield  {author} {\bibinfo {author} {\bibfnamefont {T.}~\bibnamefont {Barthel}}\ and\ \bibinfo {author} {\bibfnamefont {Y.}~\bibnamefont {Zhang}},\ }\href {https://doi.org/10.1088/1742-5468/ac8e5c} {\bibfield  {journal} {\bibinfo  {journal} {Journal of Statistical Mechanics: Theory and Experiment}\ }\textbf {\bibinfo {volume} {2022}},\ \bibinfo {pages} {113101} (\bibinfo {year} {2022})}\BibitemShut {NoStop}%
\bibitem [{\citenamefont {Mehboudi}\ and\ \citenamefont {Miller}(2022)}]{mehboudi_thermodynamic_2022}%
  \BibitemOpen
  \bibfield  {author} {\bibinfo {author} {\bibfnamefont {M.}~\bibnamefont {Mehboudi}}\ and\ \bibinfo {author} {\bibfnamefont {H.~J.~D.}\ \bibnamefont {Miller}},\ }\href {https://doi.org/10.1103/PhysRevA.105.062434} {\bibfield  {journal} {\bibinfo  {journal} {Physical Review A}\ }\textbf {\bibinfo {volume} {105}},\ \bibinfo {pages} {062434} (\bibinfo {year} {2022})}\BibitemShut {NoStop}%
\bibitem [{\citenamefont {Mehboudi}\ \emph {et~al.}(2019)\citenamefont {Mehboudi}, \citenamefont {Parrondo},\ and\ \citenamefont {Acín}}]{mehboudi_linear_2019}%
  \BibitemOpen
  \bibfield  {author} {\bibinfo {author} {\bibfnamefont {M.}~\bibnamefont {Mehboudi}}, \bibinfo {author} {\bibfnamefont {J.~M.~R.}\ \bibnamefont {Parrondo}},\ and\ \bibinfo {author} {\bibfnamefont {A.}~\bibnamefont {Acín}},\ }\href {https://doi.org/10.1088/1367-2630/ab30f4} {\bibfield  {journal} {\bibinfo  {journal} {New Journal of Physics}\ }\textbf {\bibinfo {volume} {21}},\ \bibinfo {pages} {083036} (\bibinfo {year} {2019})}\BibitemShut {NoStop}%
\bibitem [{\citenamefont {Holstein}\ and\ \citenamefont {Primakoff}(1940)}]{holstein_field_1940}%
  \BibitemOpen
  \bibfield  {author} {\bibinfo {author} {\bibfnamefont {T.}~\bibnamefont {Holstein}}\ and\ \bibinfo {author} {\bibfnamefont {H.}~\bibnamefont {Primakoff}},\ }\href {https://doi.org/10.1103/PhysRev.58.1098} {\bibfield  {journal} {\bibinfo  {journal} {Physical Review}\ }\textbf {\bibinfo {volume} {58}},\ \bibinfo {pages} {1098} (\bibinfo {year} {1940})}\BibitemShut {NoStop}%
\bibitem [{\citenamefont {Vicentini}\ \emph {et~al.}(2018)\citenamefont {Vicentini}, \citenamefont {Minganti}, \citenamefont {Rota}, \citenamefont {Orso},\ and\ \citenamefont {Ciuti}}]{vicentini_critical_2018}%
  \BibitemOpen
  \bibfield  {author} {\bibinfo {author} {\bibfnamefont {F.}~\bibnamefont {Vicentini}}, \bibinfo {author} {\bibfnamefont {F.}~\bibnamefont {Minganti}}, \bibinfo {author} {\bibfnamefont {R.}~\bibnamefont {Rota}}, \bibinfo {author} {\bibfnamefont {G.}~\bibnamefont {Orso}},\ and\ \bibinfo {author} {\bibfnamefont {C.}~\bibnamefont {Ciuti}},\ }\href {https://doi.org/10.1103/PhysRevA.97.013853} {\bibfield  {journal} {\bibinfo  {journal} {Physical Review A}\ }\textbf {\bibinfo {volume} {97}},\ \bibinfo {pages} {013853} (\bibinfo {year} {2018})}\BibitemShut {NoStop}%
\bibitem [{\citenamefont {Botet}\ \emph {et~al.}(1982)\citenamefont {Botet}, \citenamefont {Jullien},\ and\ \citenamefont {Pfeuty}}]{Botet_size_1982}%
  \BibitemOpen
  \bibfield  {author} {\bibinfo {author} {\bibfnamefont {R.}~\bibnamefont {Botet}}, \bibinfo {author} {\bibfnamefont {R.}~\bibnamefont {Jullien}},\ and\ \bibinfo {author} {\bibfnamefont {P.}~\bibnamefont {Pfeuty}},\ }\href {https://doi.org/10.1103/PhysRevLett.49.478} {\bibfield  {journal} {\bibinfo  {journal} {Phys. Rev. Lett.}\ }\textbf {\bibinfo {volume} {49}},\ \bibinfo {pages} {478} (\bibinfo {year} {1982})}\BibitemShut {NoStop}%
\end{thebibliography}%

\appendix
\section{Slow-driving expansion of the state and the thermodynamic metric}
\label{app:thermo:metric}
For sufficiently slow driving through $\mathcal{L}_{g(t)}$, away from the critical point, and starting from the NESS $\hat{\pi}_{g(0)}$, the system will remain close to $\hat{\pi}_{g(t)}$, up to a small correction. We consider the expansion
\begin{equation}
\label{eq:correction_ness}
    \hat{\rho}_t = \hat{\pi}_{g(t)} + \delta\hat{\rho}_t.
\end{equation}
Inserting this expansion into $\mathrm{d}\hat\rho/\mathrm{d}t = \mathcal{L}_{g(t)}[\hat\rho]$ and using that $\mathcal{L}_{g(t)}[\hat{\pi}_{g(t)}]=0$, we obtain a differential equation for $\delta\hat{\rho}_t$~\cite{scandi_thermodynamic_2019, mandal_analysis_2016, cavina_slow_2017}
\begin{equation}
\label{eq:time_evo_correction_ness}
    \left(\mathcal{L}_{g(t)}-\frac{\mathrm{d}}{\mathrm{d}t}\right)[\delta\hat{\rho}_t] = \frac{\mathrm{d}}{\mathrm{d}t}{\hat{\pi}_{g(t)}},
\end{equation}
which can be formally solved using the Drazin inverse~\cite{Gower1972} of the Liouvillian superoperator, $\mathcal{L}_g^+$. The Drazin inverse is defined as the unique operator satisfying the properties $\mathcal{L}_g\mathcal{L}_g^+[\hat{A}] = \mathcal{L}_g^+\mathcal{L}_g[\hat{A}] = \hat{A} - \hat{\pi}_g \Tr[\hat{A}]$, $\mathcal{L}_g^+[\hat{\pi}_g]=0$ and $\Tr[\mathcal{L}_g^+[\hat{A}]]=0$. It has an integral representation given by~\cite{Gower1972}
\begin{equation}
    \label{eq:drazin_integral}
    \mathcal{L}_g^+[\hat{A}] := \int_0^\infty \mathrm{d}\tau\, e^{\mathcal{L}_g\tau}\qty(\hat{\pi}_g \Tr[\hat{A}]-\hat{A}).
\end{equation}
By applying $\mathcal{L}_{g(t)}^+$ on both sides of Eq.~\eqref{eq:time_evo_correction_ness} and formally inverting the superoperator on the left-hand side, one obtains the geometric series expansion
\begin{equation}
    \label{eq:expansion_rho}
    \hat{\rho}_t =\hat{\pi}_{g(t)} + \sum_{n=1}^\infty\left(\mathcal{L}_{g(t)}^+\frac{\mathrm{d}}{\mathrm{d}t}\right)^n\hat{\pi}_{g(t)}.
\end{equation}
Notice that the $n^\text{th}$ term in this series contains an $n^\text{th}$-order derivative of $\hat{\pi}_{g(t)}$, and is thus proportional to $\dot g^n$, which means that Eq.~\eqref{eq:expansion_rho} can be seen as an expansion in powers of the driving speed $\dot g$.

Using the above expansion of the state, the nonadiabatic entropy production along a slow driving protocol $g(t)$ becomes~\cite{lacerda_information_2025}, to leading order in $\dot{g}$,
\begin{equation}
\Sigma_{\mathrm{na}}(\tau) = \int_0^{\tau}\mathrm{d}t\, \dot{g}^2(t)\, \zeta_{g(t)},
\end{equation}
where $\zeta_{g}$ denotes the thermodynamic metric in the space of NESS parameterized by $g$
\begin{equation}
\label{eq:metric:trace:form}
    \zeta_{g} = -\Tr\!\left[\frac{\mathrm{d}\log\hat \pi_g}{\mathrm{d}g}\,\mathcal{L}^+_g\!\left(\frac{\mathrm{d}\hat \pi_g}{\mathrm{d}g}\right)\right].
\end{equation}
Equivalently, using Eq.~\eqref{eq:drazin_integral}, this metric can be expressed as a correlation integral,
\begin{equation}
    \label{eq:green_kubo_elements}
    \zeta_{g} = \int_0^\infty \mathrm{d}\tau\,\langle \hat{F}_g, \hat{F}_g(\tau)\rangle_{\hat{\pi}_g},
\end{equation}
where the Kubo–Mori–Bogoliubov (KMB) inner product is 
\begin{equation}
\langle \hat{A}, \hat{B}\rangle_{\hat{\pi}_g} = \int_0^1 \mathrm{d}s\, \Tr\!\left[\hat{\pi}_g^s \hat{A}^\dagger \hat{\pi}_g^{1-s}\hat{B}\right],
\end{equation}
and $\hat{F}_g = \partial_g \log \hat{\pi}_g$ is the logarithmic derivative operator, evolved in the Heisenberg picture as $\hat{F}_g(\tau) = e^{\mathcal{L}_g\tau}[\hat{F}_g]$.  
The metric is closely related to the KMB quantum Fisher information (QFI),
\begin{equation}\label{eq:QFI_matrix}
    \mathcal{I}^\mathrm{KMB}_{g} = \mathrm{Tr}\left[\frac{\mathrm{d}\log \hat \pi_g}{\mathrm{d}g}\frac{\mathrm{d} \hat \pi_g}{\mathrm{d}g}\right] = \langle \hat{F}_g, \hat{F}_g\rangle_{\hat{\pi}_g},
\end{equation}
via the integral relaxation time
\begin{equation}
    \tau_g = \int_0^\infty \mathrm{d}\tau\, 
    \frac{\langle \hat{F}_g, \hat{F}_g(\tau)\rangle_{\hat{\pi}_g}}{\langle \hat{F}_g, \hat{F}_g\rangle_{\hat{\pi}_g}}.
\end{equation}
By construction, therefore,
\begin{equation}
\zeta_g = \tau_g\, \mathcal{I}^\mathrm{KMB}_{g}.
\end{equation}
\section{Scaling of the integral relaxation time with the inverse Liouvillian gap}
\label{app:tau_gap}
We now show that the integral relaxation time $\tau_g$ scales with the inverse Liouvillian gap near a dissipative critical point. 
Expanding the correlation function in the eigenbasis of the vectorized Liouvillian $\{\lambda_n, |\hat r_n\rangle, \langle \hat l_n|\}$, defined by $\mathcal{L}_g |\hat r_n\rangle = \lambda_n |\hat r_n\rangle$ and 
$\langle \hat l_n| \mathcal{L}_g = \lambda_n \langle \hat l_n|$, and biorthonormal, so that $\langle \hat l_n | \hat r_m \rangle = \delta_{nm}$, we obtain
\begin{equation}
|\hat F_g(\tau)\rangle = e^{{\mathcal L}_g \tau} |\hat F_g\rangle 
= \sum_{n} e^{\lambda_n \tau} |\hat r_n\rangle 
    \langle \hat l_n \vert \hat F_g \rangle,
\end{equation}
where $\langle \hat l_n\vert \hat F_g\rangle = \mathrm{Tr}[\hat l_n \hat F_g]$.
Substituting into the KMB correlation function yields
\begin{equation}
\langle \hat F_g, \hat F_g(\tau) \rangle_{\hat \pi_g}
= \sum_{n} e^{\lambda_n \tau}\,
   \langle \hat l_n\vert \hat F_g \rangle\,
   \langle \hat F_g, \hat r_n \rangle_{\hat \pi_g},
\end{equation}
and the integral relaxation time
\begin{equation}
\tau_g =
-\frac{\displaystyle
   \sum_{n>0} {\lambda_n}^{-1}\,
   \langle \hat l_n\vert \hat F_g \rangle\,
   \langle \hat F_g, \hat r_n \rangle_{\hat \pi_g}}
{\displaystyle
   \sum_{n>0}
   \langle \hat l_n\vert \hat F_g \rangle\,
   \langle \hat F_g, \hat r_n \rangle_{\hat \pi_g}}.
\end{equation}
Since $\mathrm{Re}(\lambda_n)<0$ for all $n>0$, the integral converges.
Close to a dissipative phase transition, the slowest mode ($n=1$) dominates the dynamics. As the Liouvillian gap is given by $\Delta_g := -\mathrm{Re}(\lambda_1)$, we obtain
\begin{equation}
\tau_g \sim \frac{1}{\Delta_g}.
\end{equation}
Importantly, this result mirrors the divergence of relaxation times in equilibrium critical phenomena. In fact, in Ref.~\cite{naze_kibblezurek_2022} the integral relaxation time near a closed-system quantum phase transition was shown to diverge with the inverse of the closing Hamiltonian gap.
\begin{figure}
    \centering
    \includegraphics[width=\linewidth]{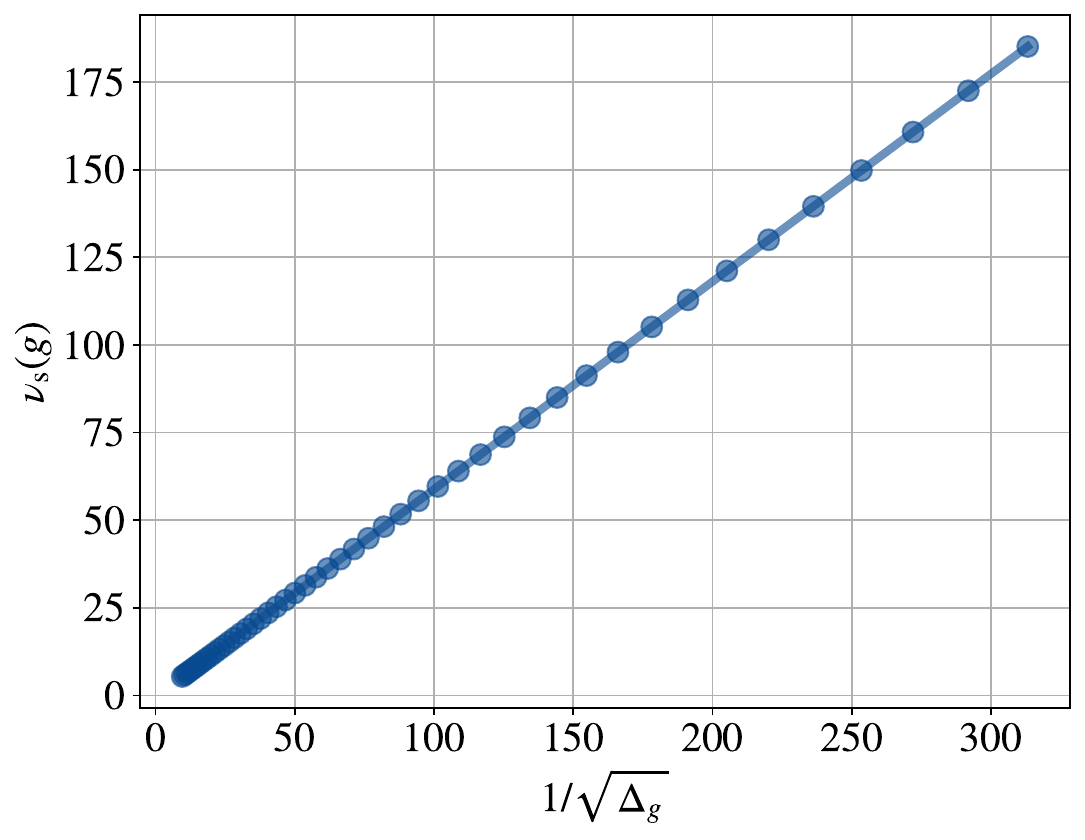}
    \caption{Scaling of the  soft mode symplectic eigenvalue with the square-root of the inverse Liouvillian gap in the dissipative Dicke model. Parameters: cavity frequency $\omega_c = 1$, atomic frequency $\omega_z = 1$, photon loss rate $\kappa = 0.2$, initial coupling strength $g_0 = g_c - 5\times 10^{-3}$, final coupling strength $g_f = g_c - 10^{-8}$ $\tau_q = 10^2\dots 10^5$.}
    \label{fig:symplectic:eigenvalue}
\end{figure}
\section{Gaussian description}
If both the Hamiltonian and the Lindblad jump operators are at most quadratic and linear in bosonic operators, respectively, the dynamics preserve Gaussianity~\cite{landi_current_2024, Serafini2023-er}.
\begin{figure*}[t]\centering\includegraphics[width=0.49\linewidth]{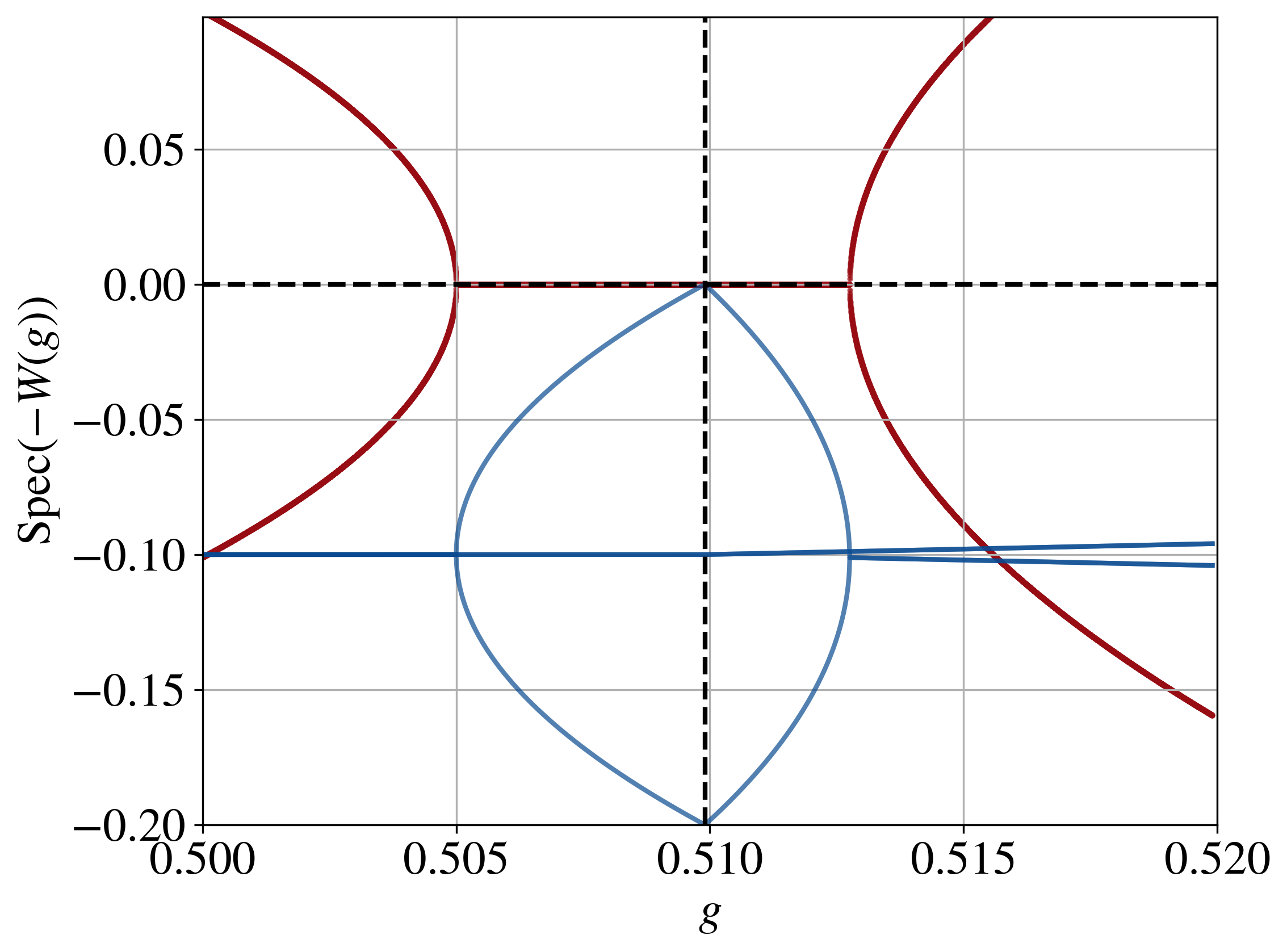}
    \includegraphics[width=0.49\linewidth]{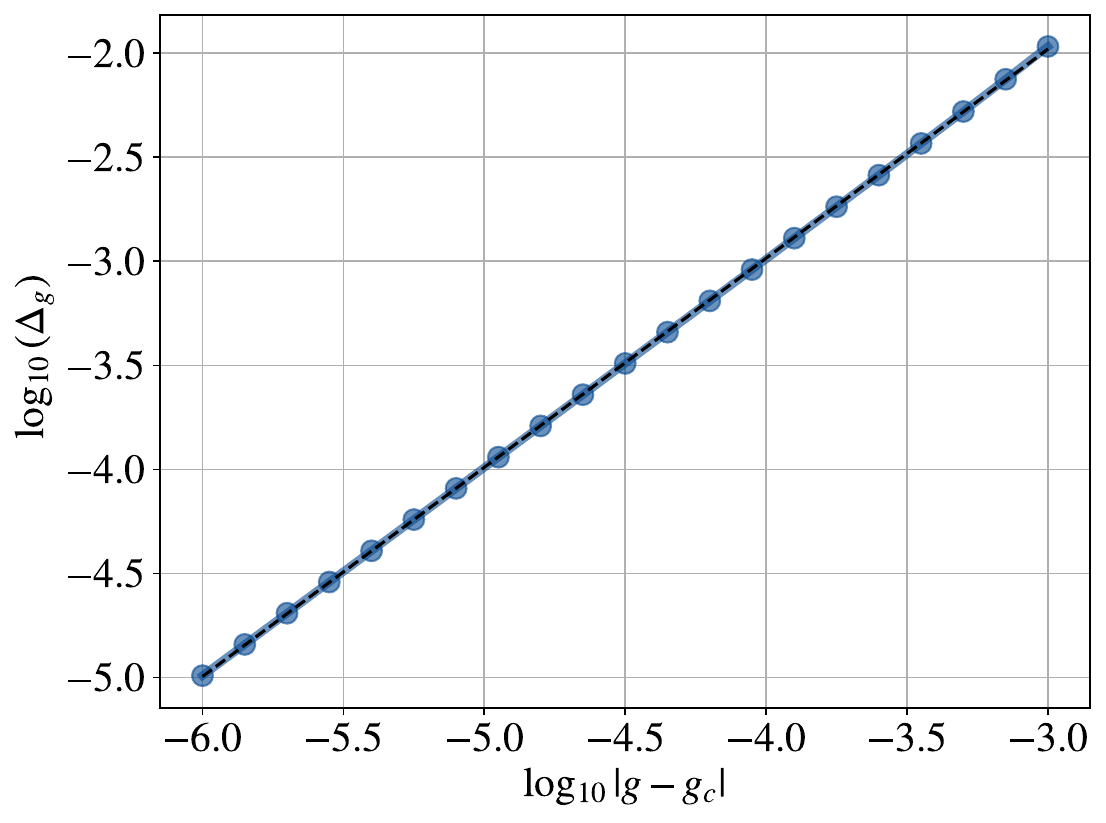}
    \caption{Left: Complex spectrum of the drift matrix $-W$ as a function of the coupling strength $g$ in the dissipative Dicke model (blue: real part, red: imaginary part). The vertical dashed black line indicates the critical point $g_c$ at which the Liouvillian gap closes. Right: Scaling of the Liouvillian gap with proximity of the coupling strength $g$ to the critical point $g_c$: $\Delta_g \sim \vert g - g_c\vert^\gamma$, in the dissipative Dicke model. From a numerical fit, we obtain $\gamma = 1.009 \pm 0.001$.   Parameters: cavity frequency $\omega_c = 1$, atomic frequency $\omega_z = 1$, photon loss rate $\kappa = 0.2$.}
    \label{fig: ODM_gap_scaling}
\end{figure*}
\subsection{General Gaussian bosonic Lindblad dynamics}
\label{app:Gaussian:dynamics}
We assume dynamics obeying a Lindblad master equation of the form
\begin{equation}
    \mathcal{L}\left[\hat\rho \right]= -i [\hat H, \hat\rho] + \sum_i \gamma_i^- \mathcal{D}[\hat b_i]\hat\rho + \gamma_i^+\mathcal{D}[\hat b_i^\dagger]\hat\rho
\end{equation}
with a bosonic quadratic Hamiltonian
\begin{equation}
\begin{split}
    \hat H = &\sum_{ij} A_{ij} \hat b_i^\dagger \hat b_j + \frac{1}{2}(B_{ij}\hat b_i^\dagger\hat  b_j^\dagger + B^*_{ij}\hat b_j \hat b_i )) 
\end{split}
\end{equation}
with $A=A^\dagger$ and $B=B^T$.
We define quadratures
\begin{equation}
    \begin{split}
       \hat  q_i = \frac{1}{\sqrt{2}} (\hat b_i + \hat b_i^\dagger)\\
        \hat p_i = \frac{i}{\sqrt{2}}(\hat b_i^\dagger - \hat b_i),
    \end{split}
\end{equation}
and construct the column vector $\vec R = (\hat q_1, \dots,\hat  q_N,\hat  p_1, \dots, \hat p_N)^T$.
The general quadratic Hamiltonian can be written in terms of quadratures as
\begin{equation}
\hat H = \frac{1}{2}  \vec R^T \cdot H_R \cdot \vec R ,
\end{equation}
where $H_R \in \mathbb{R}^{2n \times 2n}$ is a real, symmetric matrix representing the Hamiltonian coefficients, given by
\begin{equation}
\begin{split}
    H_R = &\frac{1}{2}(\mathbf{1}_2\otimes(A+A^T) + \sigma_z \otimes (B+B^*) \\
    &+ \sigma_y \otimes (A^T - A) - i\sigma_x \otimes(B-B^*))
\end{split}
\end{equation}
The covariance matrix is given by
\begin{equation}
    \Theta_{ij} = \frac{i}{2}\Omega_{ij} + \langle \hat R_j \hat R_i \rangle - \langle \hat R_i \rangle \langle \hat R_j \rangle,
\end{equation}
with symplectic form
\begin{equation}
    \Omega = (i\sigma_y)\otimes \mathbf{1}_N.
\end{equation}
The covariance matrix obeys the Lyapunov differential equation of the form
\begin{equation}
\label{eq:lyapunov:differential:equation}
    \frac{\mathrm{d}\Theta}{\mathrm{d}t} = - (W\Theta + \Theta W^\dagger) + Y
\end{equation}
with 
\begin{equation}
\begin{split}
    W &= -\Omega H_R + \frac{1}{2} \mathbf{1}_2 \otimes \Gamma,\\
    Y &= \frac{1}{2}\mathbf{1}_2 \otimes (\gamma_+ + \gamma_-).
\end{split}
\end{equation}
and $\Gamma = \gamma_- - \gamma_+$.
Consequently, the steadystate covariance matrix solves the Lyapunov equation
\begin{equation}
\label{eq:lyapunov:equation}
     W\Theta + \Theta W^\dagger = Y
\end{equation}
Further, it is often useful to express Gaussian bosonic states in the form
\begin{equation}
\label{eq:bosonic:gaussian:state}
\hat\rho = \displaystyle{\frac{e^{- \frac{1}{2} R^T \cdot M \cdot R}}{Z}}, \
\end{equation}
with $ M^\dagger = M$ (real, symmetric), and partition function $Z = \sqrt{\mathrm{det}\left(\Theta+\frac{i\Omega}{2}\right)}$.
\subsection{Constructing the Liouvillian spectrum from the drift matrix}
\label{app:Liouvillian:spectrum:drift:matrix}
The full Liouvillian $\mathcal{L}$ acts on the operator space of the density matrix, which is infinite-dimensional. 
The spectrum of $\mathcal{L}$ in such a basis includes all finite non-negative integer linear combinations (assuming the Liouvillian is diagonalisable)~\cite{prosen_quantization_2010, barthel_solving_2022}
\begin{equation}
\mathrm{Spec}(\mathcal{L}) = \left\{ -\sum_i k_i \lambda_i \middle| \lambda_i \in \mathrm{Spec}(W), k_i \in \mathbb{N}_0 \right\},
\end{equation}
where $W$ is the drift matrix in the Lyapunov equation~\eqref{eq:lyapunov:equation}.
Therefore, the Liouvillian gap is given by 
\begin{equation}
    \Delta = \min_{w\in \mathrm{Spec(W)}}\vert \mathrm{Re}(w)\vert.
\end{equation}
\subsection{Relative entropy for Gaussian bosonic states}
Given two bosonic Gaussian states, and assuming zero mean,
\begin{equation}
    \hat \rho = \frac{e^{-\frac{1}{2}R^T \cdot M \cdot R}}{Z} \quad \mathrm{and}\quad\hat \sigma = \frac{e^{-\frac{1}{2}R^T \cdot N \cdot R}}{Y},
\end{equation}
their relative entropy is given by
\begin{equation}
    D(\hat \rho \vert\vert \hat \sigma) = -\frac{1}{2}\mathrm{Tr}\left[(M-N)\Theta\right]-\log\frac{Z}{Y},
\end{equation}
where $\Theta$ denotes the covariance matrix with respect to the state $\hat \rho$.
Note that
\begin{equation}
    Z = \sqrt{\mathrm{det}\left(\Theta + \frac{i\Omega}{2}\right)}
\end{equation}
and
\begin{equation}
    M = 2i\Omega \coth^{-1}(2\Theta i \Omega), \quad \Theta = \frac{1}{2}\coth\left(\frac{i\Omega M }{2}\right)i\Omega.
\end{equation}
\begin{figure*}[t]
    \centering
    \includegraphics[width=0.49\linewidth]{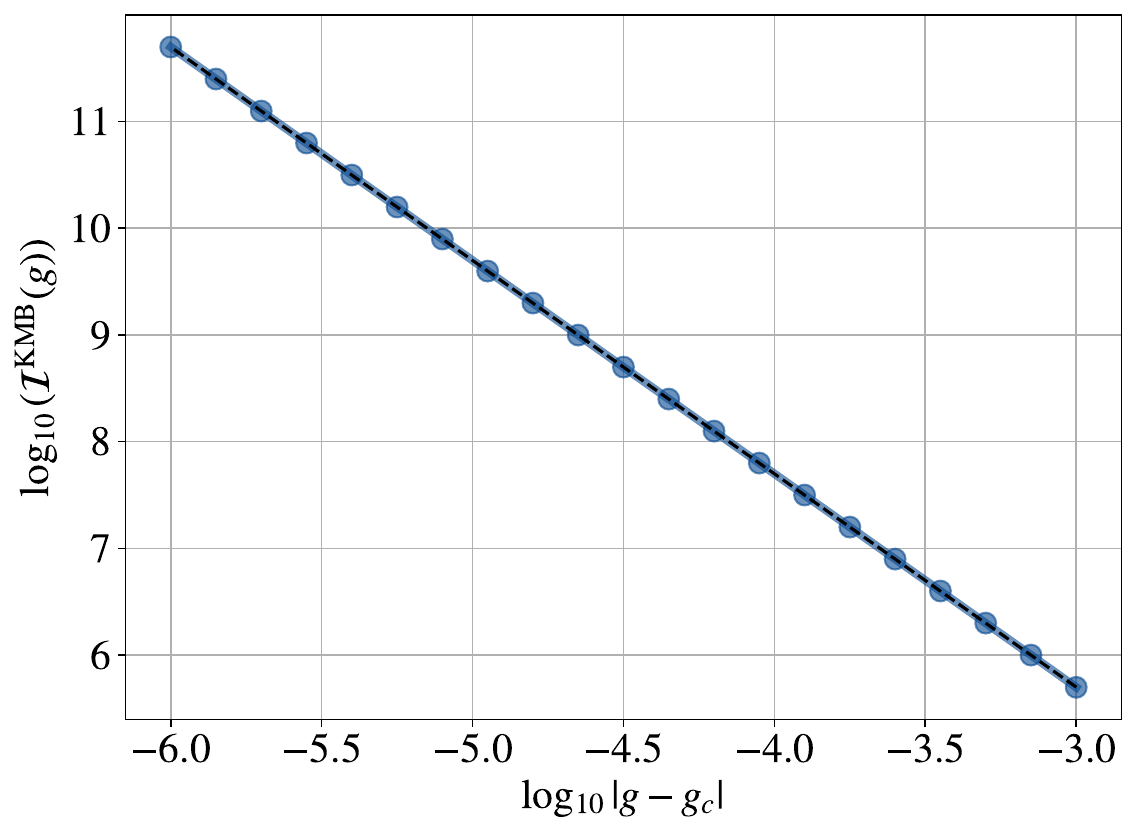}
    \includegraphics[width=0.49\linewidth]{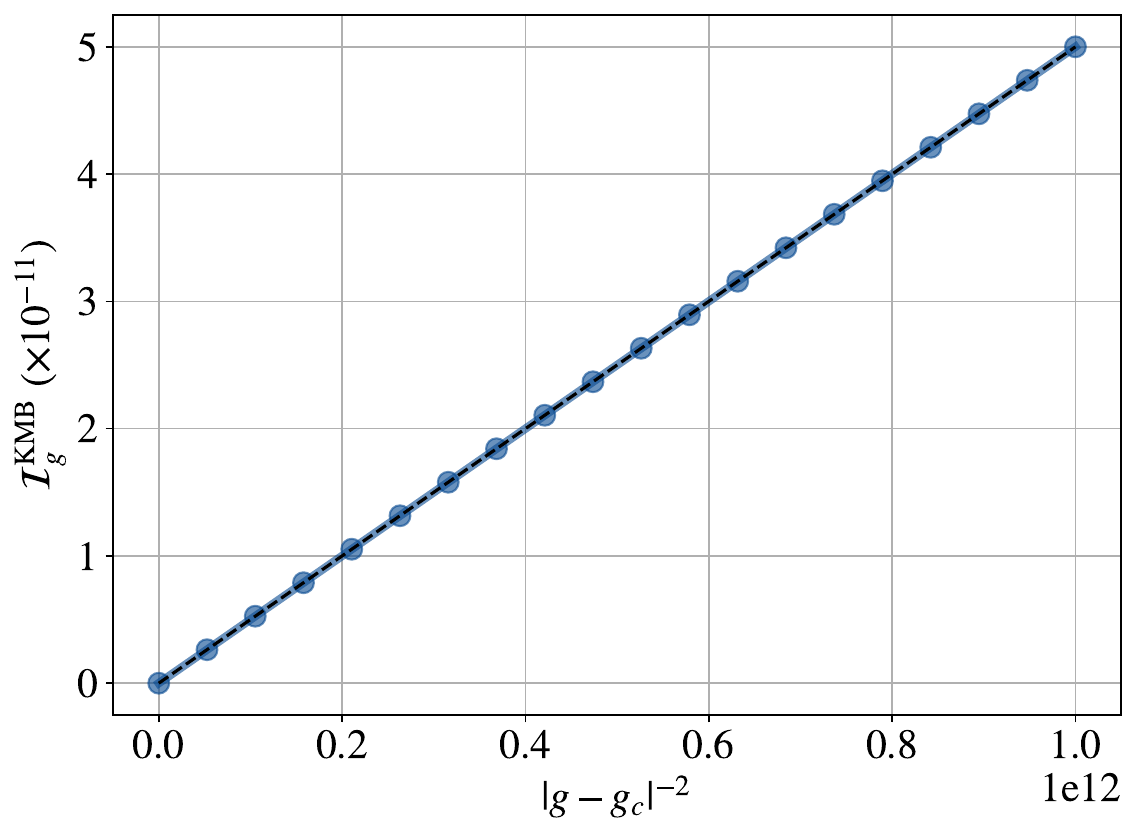}
    \caption{Left: The KMB QFI scales with a power law of the distance to the critical point $\mathcal{I}_g^\mathrm{KMB}\sim \frac{\gamma^2}{2} \vert g - g_c\vert^{-\alpha}$, with $\alpha = 2 \pm 1 \cdot 10^{-4}$, in the Gaussian open Dicke model. Right: Here we show that the slope obtained from a linear fit is $0.500001 \pm 1 \cdot 10^{-6}$, which is in excellent agreement with the predicted $\frac{\gamma^2}{2}$, with $\gamma = 1$. Parameters: cavity frequency $\omega_c = 1$, atomic frequency $\omega_z = 1$, photon loss rate $\kappa = 0.2$. }
    \label{fig:ODM:KMB:QFI}
\end{figure*}
\subsection{Thermodynamic metric for Gaussian bosonic states}
\label{app:thermodynamic:metric}
Generalizing the results of Ref.~\cite{mehboudi_thermodynamic_2022} from Gaussian bosonic thermal states to Gaussian bosonic NESS (assuming zero mean, see Eq.~\eqref{eq:bosonic:gaussian:state}), we find that the single-parameter thermodynamic metric is given by
\begin{equation}
    \zeta_{g} = \frac{1}{2}\,\mathrm{Tr}\!\left[
        \mathcal{I}(\chi_g)\, \Big(\Theta - \tfrac{i}{2}\Omega\Big)\,
        \mathcal{F}(\chi_g)\, \Big(\Theta + \tfrac{i}{2}\Omega\Big)
    \right].
\end{equation}
Here, we define $\chi_g = \partial_g M$. Furthermore,
\begin{equation}
\begin{split}
    \mathcal{I}(\chi_g) &= \int_0^1 \mathrm{d}s\, \big(e^{is\Omega M}\big)^T \chi_g e^{is\Omega M}, \\
    \mathcal{F}(\chi_g) &= \int_0^\infty d\nu\, \big(e^{\nu A}\big)^T \chi_g e^{\nu A}, \\
\end{split}
\end{equation}
with $A  = \Omega \big(H - \mathrm{Im}(CC^\dagger)\big)$, where the the matrix $C= (\vec c_1^T, \dots, \vec c_{N_L}^T)^T$ encodes the jump rates of the associated linear Lindblad jump operators $\hat L_i = \vec c_i^T \cdot \vec{R}$.
In the following, we discuss how this metric can be evaluated efficiently numerically. 
\subsection{Unitary contribution}
We first evaluate the integral in $\mathcal{I}(\chi_g)$.
To this end, we define
\begin{equation}
A := (i\Omega M)^T, \qquad B := i\Omega M.
\end{equation}
Diagonalizing yields
\begin{equation}
A = U_A D_A U_A^{-1}, \qquad B = U_B D_B U_B^{-1},
\end{equation}
with $D_A, D_B$ diagonal, and we can write
\begin{equation}
e^{sA} = U_A e^{sD_A} U_A^{-1}, \qquad e^{sB} = U_B e^{sD_B} U_B^{-1}.
\end{equation}
Substituting into $\mathcal{I}(\chi_g)$, we find
\begin{align*}
    \mathcal{I}(\chi_g) = \int_0^1 \mathrm{d}s\, e^{sA} \chi_g e^{sB}
    &= U_A \Big[ \int_0^1 \mathrm{d}s\, e^{sD_A} {\tilde{\chi}_g} e^{sD_B} \Big] U_B^{-1}, \\
    {\tilde{\chi}_g} & = U_A^{-1} \chi_g U_B.
\end{align*}
Let $D_A = \mathrm{diag}(\lambda_1,\dots,\lambda_{2n})$ and $D_B = \mathrm{diag}(\mu_1,\dots,\mu_{2n})$. Then
\begin{equation}
\big[e^{s D_A} {\tilde{\chi}_g} e^{s D_B}\big]_{ij}
  = (\tilde{\chi}_g)_{ij} e^{s(\lambda_i + \mu_j)}.
\end{equation}
Hence
\begin{equation}
\int_0^1 \mathrm{d}s\, e^{s(\lambda_i+\mu_j)} = 
\begin{cases}
\dfrac{e^{\lambda_i+\mu_j} - 1}{\lambda_i+\mu_j}, & \lambda_i+\mu_j \neq 0, \\
1, & \lambda_i+\mu_j = 0.
\end{cases}
\end{equation}
Define
\begin{equation}
J_{ij} = (\tilde{\chi}_g)_{ij} \, f(\lambda_i+\mu_j), \qquad
f(z) = \begin{cases}
\dfrac{e^z - 1}{z}, & z \neq 0, \\
1, & z = 0.
\end{cases}
\end{equation}
Then
\begin{equation}
\int_0^1 \mathrm{d}s\, e^{s (i\Omega M)^T} \chi_g e^{s i\Omega M}
   = U_A J U_B^{-1}.
\end{equation}
\subsection{Dissipative-like contribution}
For the dissipative-like term we evaluate
\begin{equation}
    \mathcal{F}(\chi_g)= \int_0^\infty d\nu\, e^{\nu A^T} \chi_g e^{\nu A}.
    \label{eq:integral_dissipative}
\end{equation}
This integral is the solution of the continuous Lyapunov equation
\begin{equation}
    A^T Y + Y A = -\chi_g,
\end{equation}
which can be solved efficiently using standard numerical routines.
\subsection{Kubo-Mori-Bogoliubov Quantum Fisher information for Gaussian bosonic states}
\label{app:QFI_KMB_Gaussian}
The Kubo-Mori-Bogoliubov (KMB) quantum Fisher information (QFI) is defined as
\begin{equation}
    \mathcal{I}_g^\mathrm{KMB} = \int_0^1\mathrm{d}s\, \mathrm{Tr}\Big[ \hat \rho^s (\partial_g \log \hat \rho) \, \hat \rho^{1-s} (\partial_g \log \hat \rho) \Big].
\end{equation}
Performing an analogous calculation to that presented in Appendix~\ref{app:thermodynamic:metric} and Refs.~\cite{mehboudi_linear_2019, mehboudi_thermodynamic_2022}, we find that for Gaussian bosonic states with zero mean (see Eq.~\eqref{eq:bosonic:gaussian:state}) , the KMB QFI is given by
\begin{equation}
\begin{split}
    \mathcal{I}_g^\mathrm{KMB} 
    =& \frac12 \int_0^1\mathrm{d}s\, \mathrm{Tr}[ \chi_g \check{\Theta} S^T(s) \chi_g S(s) \check{\Theta}^T ],
\end{split}
\end{equation}
with $S(s) = e^{si\Omega M}$, $\chi_g = \partial_g M$ and $ \check{\Theta} = \Theta + i\Omega/2$.
\subsection{Critical scaling of the symplectic eigenvalue for Gaussian bosonic states}
\label{app:scaling:soft:symplectic_eigenvalue}
We consider a Gaussian bosonic system whose covariance matrix evolves under a Lyapunov differential equation (see Eq.~\eqref{eq:lyapunov:differential:equation}), so that the steady-state covariance matrix solves the Lyapunov equation (see Eq.~\eqref{eq:lyapunov:equation}).
In this section, our aim is to understand the behaviour of the soft mode symplectic eigenvalue of the steadystate covariance matrix in the vicinity of a critical point $g_c$ in a second-order DQPT.
Let $\lambda_s$ denote the eigenvalue of $W$, assumed to be diagonalisable, with smallest absolute real part (Liouvillian gap), and
\begin{equation}
W v_s = \lambda_s v_s, \qquad w_s^\dagger W = \lambda_s w_s^\dagger,
\end{equation}
where $v_s$ is the right eigenvector and $w_s^\dagger$ is the left eigenvector, such that
\begin{equation}
w_s^\dagger v_s = 1 .
\label{eq:biorth}
\end{equation}
We define the projection of $\Theta$ onto the subspace of the slowest-decaying (soft) mode as
\begin{equation}
\theta_s = w_s^\dagger \, \Theta \, w_s .
\label{eq:theta_s}
\end{equation}
Multiplying Eq.~\eqref{eq:lyapunov:equation} from the left by $w_s^\dagger$ and from the right by $w_s$ gives
\begin{equation}
(w_s^\dagger W) \Theta w_s + w_s^\dagger \Theta (W^\dagger w_s) = w_s^\dagger Y w_s .
\end{equation}
Using the eigenvector relations and \eqref{eq:biorth}, we obtain the equation
\begin{equation}
\lambda_s \, \theta_s + \lambda_s^* \, \theta_s = y_s ,
\label{eq:scalar_proj}
\end{equation}
where
\begin{equation}
y_s = w_s^\dagger Y w_s,
\end{equation}
which we assume remains finite and non-zero across the transition.
Thus,
\begin{equation}
\theta_s = \frac{y_s}{\lambda_s + \lambda_s^*} = \frac{y_s}{2 \, \mathrm{Re} \, \lambda_s}.
\label{eq:theta_s_final}
\end{equation}
To understand the structure of divergences, recall that the Lyapunov equation admits the integral solution
\begin{equation}
\Theta = \int_0^\infty e^{tW} Y e^{tW^\dagger} \,\mathrm{d}t
= \sum_{i,j} \frac{v_i \, (w_i^\dagger Y  w_j) \,  v_j^\dagger}{\lambda_i+\lambda_j^*},
\end{equation}
using $e^{tW} = \sum_i e^{t\lambda_i}  v_i w_i^\dagger$ and $e^{tW^\dagger} = \sum_j e^{t\lambda_j^*}   w_j  v_j^\dagger$.
As $\Delta_g=\mathrm{Re}\,\lambda_s \to 0^+$, the dominant term is the contribution
\begin{equation}
\Theta = \frac{y_s}{\lambda_s+\lambda_s^*} \, v_s v_s^\dagger + \Theta_{\mathrm{reg}},
\end{equation}
where $\Theta_{\mathrm{reg}}$ remains finite at the transition. 
The divergent part is rank-one. Hence in the two-dimensional canonical subspace of the soft mode, $\Theta_s$ has one eigenvalue $\propto \Delta_g^{-1}$ and the other $\propto \mathcal{O}(1)$.
The covariance matrix of the slowest decaying mode, in canonical coordinates, can be written as
\begin{equation}
\Theta_{s} = 
\begin{pmatrix}
\langle X_s^2 \rangle & \tfrac{1}{2} \langle X_s P_s + P_s X_s \rangle \\
\tfrac{1}{2} \langle P_s X_s + X_s P_s \rangle & \langle P_s^2 \rangle
\end{pmatrix},
\end{equation}
where $(X_s,P_s)$ are the associated quadratures, chosen such that they satisfy canonical commutation relations. 
The symplectic eigenvalue of this matrix is
\begin{equation}
\nu_{s} = \sqrt{\det \Theta_{s}}.
\label{eq:nu_soft}
\end{equation}
From the rank-one structure above, we have
\begin{equation}
\det \Theta_{s} \sim \Delta_g^{-1},
\qquad 
\nu_{s} \sim \Delta_g^{-1/2}.
\end{equation}
The scaling behaviour of the soft mode symplectic eigenvalue is confirmed in the Gaussian driven-dissipative Dicke model, see Fig.~\ref{fig:symplectic:eigenvalue}
\label{app:finite:size:scaling}
\begin{figure}
    \centering
    \includegraphics[width=0.99\linewidth]{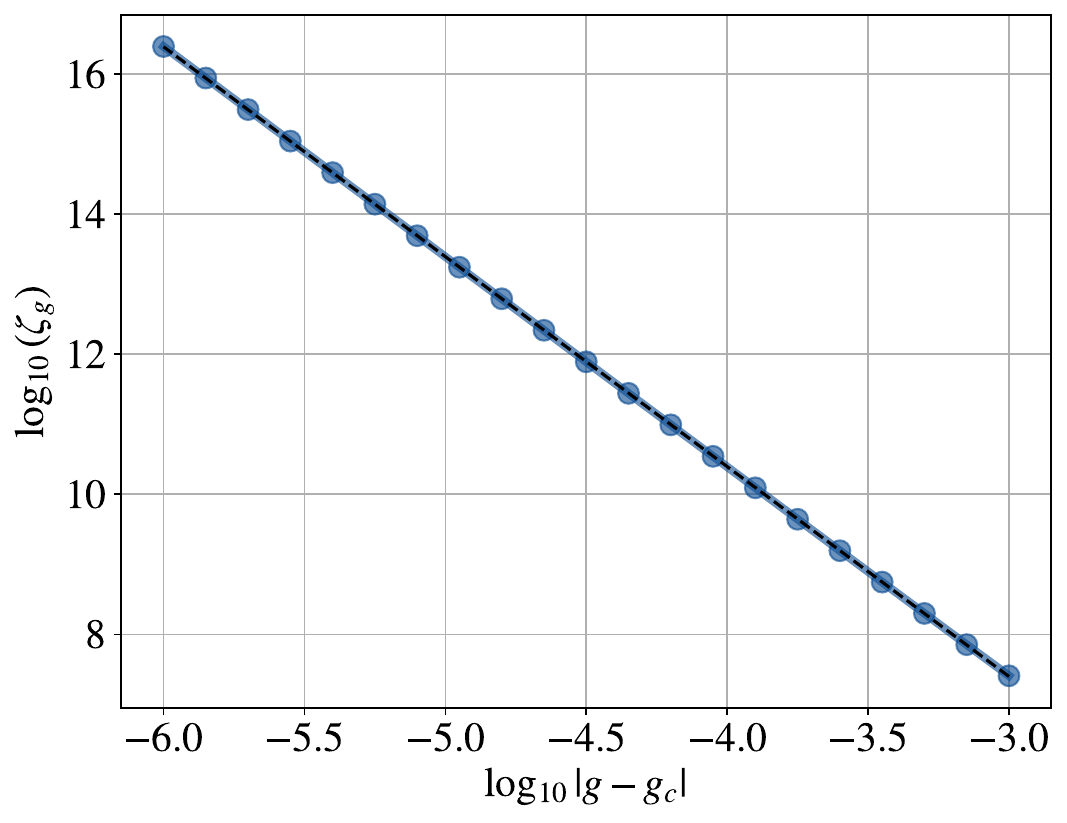}
    \caption{Scaling of the thermodynamic metric $\zeta$ near the critical point: $\zeta \sim \vert g - g_c\vert^x$, in the dissipative Dicke model. We obtain $x = -2.9953 \pm  0.0009$ via a numerical fit, which is in excellent agreement with our theoretical prediction $x = -(\alpha + \gamma)$, with $\alpha = 2$ and $\gamma = 1$. Parameters: cavity frequency $\omega_c = 1$, atomic frequency $\omega_z = 1$, photon loss rate $\kappa = 0.2$. }
    \label{fig:dicke:metric}
\end{figure}
\subsection{Scaling of the Kubo-Mori-Bogoliubov Quantum Fisher information for Gaussian bosonic states}
\label{app:KMB-scaling}
Here we show that the Kubo-Mori-Bogoliubov (KMB) quantum Fisher information (QFI) diverges as $\mathcal{I}_g^\mathrm{KMB}\sim\frac{\gamma^2}{2}|g-g_c|^{-2}$ near a critical point for bosonic Gaussian NESS with a single soft mode. 
In Appendix~\ref{app:QFI_KMB_Gaussian}, we have shown that for zero mean Gaussian states, the KMB QFI can be expressed as
\begin{equation}
\begin{split}
    \mathcal{I}_g^\mathrm{KMB} 
    &= \frac12 \int_0^1\mathrm{d}s\, \mathrm{Tr}[ A \check{\Theta} S^T(s) A S(s) \check{\Theta}^T ],
\end{split}
\end{equation}
with $S(s) = e^{si\Omega M}$.
Let us first express $X = i\Omega M$ in its biorthonormal basis
\begin{equation}
    X = \sum_i \lambda_i v_i w_i^T, \qquad w_j^T v_i = \delta_{ij},
\end{equation}
so that
\begin{equation}
    S(s) = \sum_j e^{\lambda_j s} v_j w_j^T, \qquad S^T(s)=\sum_j e^{\lambda_j s} w_j v_j^T.
\end{equation}
The KMB QFI then reads
\begin{equation}
\label{eq:I_KMB_full_components}
    \begin{split}
        \mathcal{I}^\mathrm{KMB}_g &= \frac{1}{2}\sum_{ij} \mathrm{Tr}\left[A \check \Theta \int_0^1 e^{(\lambda_i + \lambda_j)s}\,ds w_iv_i^T A v_j w_j^T \check \Theta^T\right]\\
        &= \frac{1}{2}\sum_{ij} \frac{e^{\lambda_i + \lambda_j}-1}{\lambda_i + \lambda_j} \mathrm{Tr}\left[A \check \Theta  w_iv_i^T A v_j w_j^T \check \Theta^T\right].
    \end{split}
\end{equation}
Equation~\eqref{eq:I_KMB_full_components} contains the factors $\check{\Theta} = \Theta + i\Omega/2$ and $\check{\Theta}^T = \Theta - i\Omega/2$. We expect the soft mode component to the $\Theta \Theta$-contribution to dominate near the critical point, since, as we have shown in Appendix~\ref{app:scaling:soft:symplectic_eigenvalue}, $\nu_s\sim 1/\sqrt{\Delta_g}$ and $\Delta_g \to 0$.
We further use that
\begin{equation}
\begin{split}
    A &= \partial_g M = i\Omega \partial_g X \\
    &= \sum_j \underbrace{i \Omega\partial_g \lambda_j v_jw_j^T}_{A_j^\lambda}  +\underbrace{i \Omega \lambda_j \partial_g(v_j w_j^T)}_{A_j^P}\\
    & = \sum_j A_j^\lambda + {A_j^P},
\end{split}
\end{equation}
so that $A$ can be split into components involving the derivatives of the symplectic eigenvalues ($A_j^\lambda$) the projectors ($A_j^P$), respectively.
We now perform the explicit computation of the term involving the $g-$derivatives of $\lambda_i$, labelled with $\Theta\Theta\lambda\lambda$. 
We express $\Theta = (\Theta i\Omega) i\Omega = \sum_j \nu_j v_j w_j^T i \Omega$, and find
\begin{widetext}
    \begin{equation*}
    \begin{split}
    \mathcal{I}^\mathrm{KMB, \Theta\Theta\lambda\lambda}_g &= \frac{1}{2}\sum_{ijklmn} \frac{e^{\lambda_i + \lambda_j}-1}{\lambda_i + \lambda_j} \mathrm{Tr}\left[\underbrace{i \Omega \partial_g \lambda_k v_k w_k^T}_{A_k^\lambda}\underbrace{\nu_l v_l w_l^T i\Omega}_{\Theta_l}  w_iv_i^T \underbrace{i\Omega \partial_g \lambda_m v_m w_m^T}_{A^\lambda_m} v_j w_j^T \underbrace{\nu_n v_n w_n^T i \Omega}_{\Theta_n} \right]\\
        &= \frac{1}{2}\sum_{ij} \frac{e^{\lambda_i + \lambda_j}-1}{\lambda_i + \lambda_j} (\partial_g \lambda_j \nu_j)^2\,\, w_j^T i \Omega w_i \,\,v_i^T i \Omega v_j,
    \end{split}
\end{equation*}
\end{widetext}
where have used that $(i\Omega)^2 = 1$, the cyclic property of the trace as well as the biorthonormality condition.
Since $X = 2 \coth^{-1}(2\Theta i \Omega)$, $\Theta i \Omega$ shares an eigenbasis with $X$ and their respective eigenvalues are related via 
$\lambda_j = 2 \coth^{-1}(2\nu_j)$,
and 
\begin{equation}
    \partial_g\lambda_j = -\frac{4\partial_g \nu_j}{4 \nu_j^2 -1}\overset{g\to g_c}{\longrightarrow}-\frac{\partial_g \nu_j}{\nu_j^2 } .
\end{equation}
From our analysis of the soft mode in Appendix~\ref{app:scaling:soft:symplectic_eigenvalue}, we know that in the vicinity of the critical point, the soft mode symplectic eigenvalue of the covariance matrix diverges as $\nu_s \sim 1/\sqrt{\Delta_g}$. Therefore, the soft mode symplectic eigenvalue of $M$ $\lambda_s\sim 1/\nu_s$ vanishes.
Further, the factors $w_j^T i \Omega w_i $ and $v_i^T i \Omega v_j$ are nonzero only for the labels $i$ and $j$ referring to symplectic eigenvalues of opposite sign. Therefore, there are two nonzero identical contributions to the sum, relevant to the scaling $(s, -s)$ and $(-s, s)$.
Further, we note that the prefactor $\frac{e^{\lambda_i + \lambda_j}-1}{\lambda_i + \lambda_j} $ reduces to 1, for the dominant soft mode contribution, since $\lambda_s \to 0$, and find
\begin{equation}
\mathcal{I}^\mathrm{KMB, \Theta\Theta\lambda\lambda}_g  \sim (\partial_g \lambda_j )^2\nu_j^2\approx\left(\frac{\partial_g \nu_s}{\nu_s}\right)^2\sim \left(\frac{\gamma}{2}\right)^2\vert g-g_c\vert^{-2},
\end{equation}
Similarly, we find for the contribution involving the projector derivatives
\begin{widetext}
    \begin{equation*}
    \begin{split}
    \mathcal{I}^\mathrm{KMB, \Theta\Theta P P }_g &= \frac{1}{2}\sum_{ijklmn} \frac{e^{\lambda_i + \lambda_j}-1}{\lambda_i + \lambda_j} \mathrm{Tr}\left[\underbrace{i \Omega \lambda_k \partial_g (v_k w_k^T)}_{A_k^P}\underbrace{\nu_l v_l w_l^T i\Omega}_{\Theta_l}  w_iv_i^T \underbrace{i\Omega  \lambda_m \partial_g( v_m w_m^T)}_{A^P_m} v_j w_j^T \underbrace{\nu_n v_n w_n^T i \Omega}_{\Theta_n} \right]\\
        &\sim (\partial_g \lambda_j )^2\nu_j^2\sim \left(\frac{\gamma}{2}\right)^2\vert g-g_c\vert^{-2}.
    \end{split}
\end{equation*}
\end{widetext}
All other terms can be computed in an analogous fashion and can be shown to be subleading or vanishing by the biorthonormality condition.
Combining these findings, we obtain that the dominant scaling of the KMB QFI is
\begin{equation}
    \mathcal{I}^\mathrm{KMB}_g  \sim \frac{\gamma^2}{2}\vert g-g_c\vert^{-2},
\end{equation}
so that the KMB QFI critical exponent is given by $\alpha =2$. Note that the prefactor $\frac{\gamma^2}{2}$ depends on the gap exponent $\gamma$.
\section{Driven-dissipative Dicke model in the thermodynamic limit}
\label{app:Dicke:model}
The quantum Dicke model~\cite{dicke_coherence_1954, hepp_equilibrium_1973, wang_phase_1973, hioe_phase_1973, klinder_dynamical_2015, lang_critical_2016, paz_driven-dissipative_2022, dimer_proposed_2007, nagy_critical_2011, oztop_excitations_2012} describes the interaction between a single-mode bosonic field and a large ensemble of two-level atoms (spins). The Hamiltonian (in units where $\hbar = 1$) reads
\begin{equation}
\hat H = \omega_c \hat a^\dagger \hat a + \omega_z \hat J_z + \frac{g}{\sqrt{N}} (\hat a + \hat a^\dagger)(\hat J_+ + \hat J_-),
\label{eq:DickeHamiltonian}
\end{equation}
where $\hat a, \hat a^\dagger$ are the annihilation and creation operators of the cavity mode, $\hat J_z, \hat J_\pm$ are collective spin operators for $N$ two-level atoms, $\omega_c$ is the cavity photon frequency, $\omega_z$ is the atomic transition frequency, and $g$ is the atom-light coupling strength.
In the thermodynamic limit $N \to \infty$, we assume that the system stays close to the collective spin ground state. We apply the Holstein-Primakoff transformation~\cite{holstein_field_1940, dimer_proposed_2007}
\begin{align}
\hat J_+ &= \hat b^\dagger \sqrt{2J - \hat b^\dagger \hat b} \approx \sqrt{2J}\, \hat b^\dagger, \\
\hat J_- &= \sqrt{2J - \hat b^\dagger \hat b}\, \hat b \approx \sqrt{2J}\, \hat b, \\
\hat J_z &= \hat b^\dagger \hat b - J,
\end{align}
where $J = N/2$ and $\hat b$ is a bosonic operator with $[\hat b, \hat b^\dagger] = 1$.
Substituting into Eq.~\eqref{eq:DickeHamiltonian} and taking the large-$N$ limit, the Hamiltonian in the normal phase becomes ($g <g_c$)
\begin{align}
\hat H_N &\approx \omega_c \hat a^\dagger \hat a + \omega_z \hat b^\dagger \hat b  + g (\hat a + \hat a^\dagger)(\hat b + \hat b^\dagger), \label{eq:HPH}
\end{align}
which is a quadratic Hamiltonian in the bosonic modes $\hat a$ and $\hat b$.
To include dissipation, we add a Lindblad term modeling photon loss: 
\begin{equation}
\dot{\hat\rho} = -i[\hat H, \hat\rho] + \kappa \mathcal{D} [\hat a]\hat\rho, 
\end{equation}
where $\mathcal{D}[\hat L]\hat\rho = \hat L\hat\rho \hat L^\dagger - \frac{1}{2}\{\hat L^\dagger \hat L, \hat\rho\}$, and $\kappa$ is the cavity decay rate. 
For the driven-dissipative Dicke model, we find the matrices (see Appendix~\ref{app:Gaussian:dynamics})
\begin{equation}
    A_N = \left(\begin{matrix}
         \omega_c & g \\
    g & \omega_z
    \end{matrix}\right)
\end{equation}
and
\begin{equation}
    B_N = \left(\begin{matrix}
         0 & g \\
    g & 0
    \end{matrix}\right)
\end{equation}
for the normal phase Hamiltonian.
Further,
\begin{equation}
    \begin{split}
    \Gamma &=
\begin{pmatrix}
2\kappa & 0  \\
0 & 0
\end{pmatrix}.
    \end{split}
\end{equation}
We can obtain the critical point by first setting $\mathrm{det}(W)=0$, and find~\cite{kirton_introduction_2019} 
\begin{equation}
    g_c = \frac{1}{2}\sqrt{\frac{\omega_c^2 + \kappa^2}{\omega_c} \omega_z}.
\end{equation}
\subsection{Critical scaling of the Liouvillian gap}
The Liouvillian gap can be obtained from the spectrum of the drift matrix $W$, as described in Appendix~\ref{app:Liouvillian:spectrum:drift:matrix}.
Fig.~\ref{fig: ODM_gap_scaling} shows the complex spectrum of $W$ as a function of the coupling strength $g$ as well as the Liouvillian gap $\Delta_g$ near the critical point. From a numerical fit, we find $\gamma = 1.009 \pm 0.001$ in the normal phase. We thus conclude $\gamma\approx 1$. This is in agreement with Ref.~\cite{torre_keldysh_2013}.
\subsection{Critical scaling of KMB QFI}
For the driven-dissipative Dicke model, we find that the KMB QFI scales with the distance to the critical point as $ I^{\mathrm{KMB}}_g \sim\frac{\gamma^2}{2} \vert g-g_c\vert^{-2}$, as shown in Fig.~\ref{fig:ODM:KMB:QFI}.
This is in excellent agreement with our analytical scaling prediction (see Appendix~\ref{app:KMB-scaling}).
\subsection{Critical scaling of the thermodynamic metric}
For the dissipative Dicke model, we find that the thermodynamic metric scales as $\zeta(g)\sim \vert g-g_c\vert^{-3}$, as shown in Fig.~\ref{fig:dicke:metric}. This numerical finding is in excellent agreement with the analytical scaling predictions (see Appendix~\ref{app:KMB-scaling} and Ref.~\cite{torre_keldysh_2013}).
\begin{figure*}[t]
    \centering
    \includegraphics[width=0.49\linewidth]{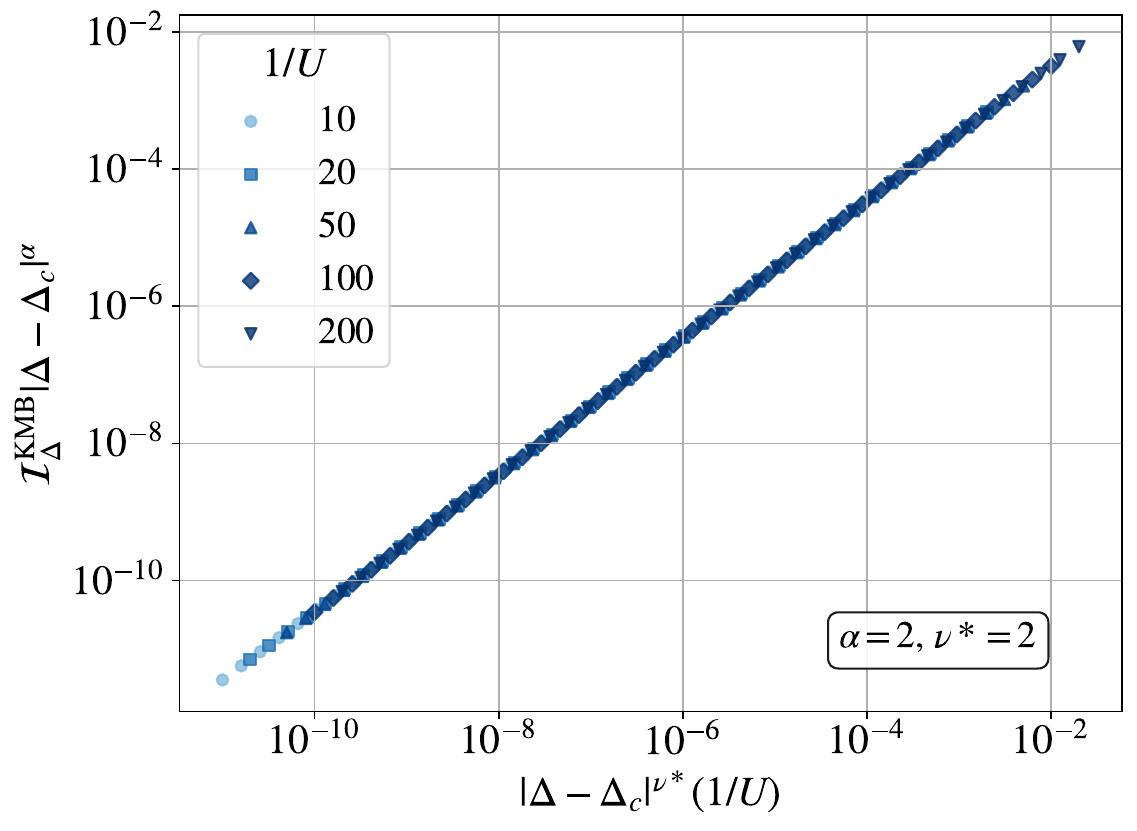}
    \includegraphics[width=0.49\linewidth]{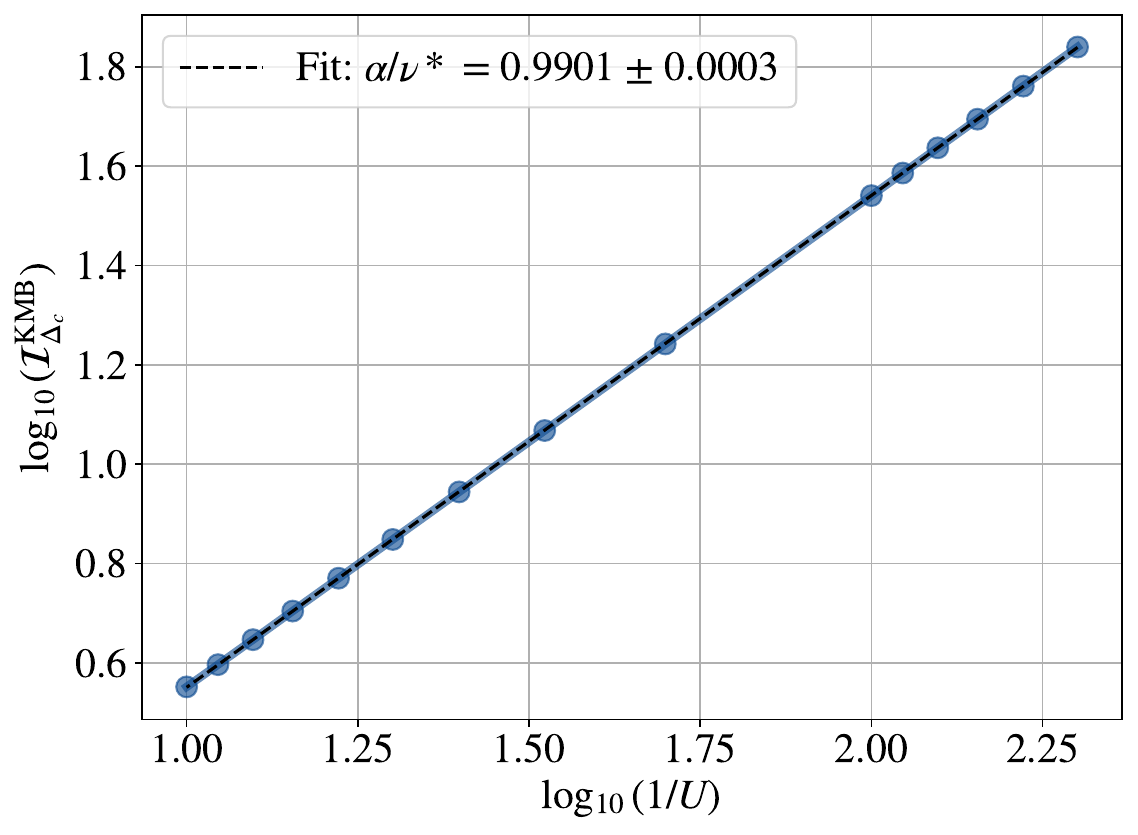}
    \caption{Finite-size scaling analysis of the KMB QFI in the dissipative Kerr model as a function of the distance to the critical point (left panel) and at the critical point (right panel). Here $1/U\to\infty$ corresponds to the thermodynamic limit~\cite{vicentini_critical_2018}. Parameters: parametric drive amplitude $G = 1.5$, photon loss rate $\kappa = 1$. }
\label{fig:kerr:finite:size:IKMB}
\end{figure*}
\section{Open Kerr parametric oscillator}
\label{app:Kerr:model}
In this section, we provide a discussion of the single-mode Kerr parametric oscillator, a nonlinear oscillator subject to two-photon driving and single photon loss, near a dissipative quantum phase transition.
The Hamiltonian of a Kerr parametric oscillator in the rotating frame reads~\cite{zhang_driven-dissipative_2021}
\begin{equation}
\hat H = -\Delta \, \hat a^\dagger \hat a + \frac{U}{2} \hat a^{\dagger 2} \hat a^2 +  \frac{1}{2}(G\hat a^{\dagger 2} + G^*\hat a^2),
\end{equation}
where $\hat a$ and $\hat a^\dagger$ are the bosonic ladder operators, $\Delta$ is the detuning between the driving and cavity frequency, $U$ the Kerr nonlinearity, and $G$ the parametric drive amplitude.  Here $1/U\to\infty$ corresponds to the thermodynamic limit~\cite{vicentini_critical_2018}.
The system is coupled to a Markovian environment with one-photon loss at rate $\kappa$, described by the Lindblad master equation
\begin{equation}
\frac{\mathrm{d}}{\mathrm{d}t} \hat \rho = -i[\hat H, \hat \rho] + \kappa \mathcal{D}[\hat a] \hat \rho, \quad
\mathcal{D}[\hat a] \hat\rho = \hat a \hat \rho \hat a^\dagger - \frac{1}{2} \{\hat a^\dagger \hat a, \hat \rho\}.
\end{equation}
\subsection{Finite-size scaling}
In infinitely coordinated models, or models without spatial degrees of freedom, the scaling analysis relies on the so-called coherence number, instead of the correlation length, which is assumed to diverge in the thermodynamic limit, as~\cite{Botet_size_1982, Hwang_dissipative_2018}
\begin{equation}
    N_c \sim \vert g-g_c \vert^{-\nu^*}.
\end{equation}
We are interested in the finite-size scaling behaviour of a quantity that diverges in the thermodynamic limit with proximity to the critical point as 
\begin{equation}
    A_\infty\sim \vert g-g_c\vert^{-\alpha_A}.
\end{equation}
The scaling hypothesis assumes the existence of a regular function $F_A$, s.t. near $g_c$ and for large N
\begin{equation}
    A_N\sim \vert g-g_c\vert^{-\alpha_A} F_A\left(\frac{N}{N_c}\right)
\end{equation}
Therefore, we expect a collapse of the curves when plotting
$A_N \vert g-g_c\vert^{\alpha_A}\quad \mathrm{vs.} \quad N \vert g-g_c\vert^{\nu^*}$ and further $A_N(g_c)\sim N^{\alpha_A/\nu *}$.
Interestingly, $\nu^*$ is related to the mean field $\nu_\mathrm{MF}$ and the upper critical dimension $d_c$ of a corresponding short-range mean field model in dimension $d = d_c$ via $\nu^* = \nu_\mathrm{MF} d_c$.
In Fig.~\ref{fig:kerr:finite:size:IKMB} we present the finite-size scaling results for the KMB QFI of the dissipative Kerr model. The numerical findings confirm $\alpha = 2$ and are in agreement with $\nu_\mathrm{MF} = 1/2$ and $d_c = 4$ of the mean-field Ising universality class with $\mathbb{Z}_2$-symmetry in the critical dimension. Further, our results agree with similar analysis of models in the same universality class~\cite{Hwang_dissipative_2018, Liu_universal_2025}.
\end{document}